\documentclass{aa}  
\usepackage[switch]{lineno} 

\usepackage{graphicx}
\usepackage{txfonts}
 \usepackage{hyperref}
 \usepackage{subcaption}
 \usepackage{xcolor}
\begin{document} 
    \titlerunning{Gaia XP in Red Giant Asteroseismology}
    \authorrunning{Barman et al.}

   \title{Potential of Gaia XP Spectra in Red Giant Star Asteroseismology: A Deep-Learning Approach}


   \author{Rajarshi Barman
          \inst{1,2},
          Shatanik Bhattacharya\inst{2},
          Shravan M. Hanasoge\inst{2}
          \and
          Siddharth Dhanpal\inst{2},
          }

   \institute{Department of Physics, Indian Institute of Science,
             Bengaluru, India \\
              \email{rajarshiwork123@gmail.com}
         \and
             Department of Astronomy and Astrophysics, Tata Institute of Fundamental Research, Mumbai, India\\
             }

   \date{Received xxx; accepted xxx}

 
  \abstract
   {Red giants are key tracers of stellar evolution and Galactic structure, and their asteroseismic properties — particularly the large frequency separation ($\Delta\nu$), the frequency of maximum oscillation power ($\nu_{\text{max}}$), and the dipole-mode period spacing ($\Delta\Pi_{1}$) which — provide direct insight into their internal structure, masses, and evolutionary states. Until now, seismic inferences on large stellar samples have relied primarily on high-quality light curves from missions such as \textit{Kepler} and \textit{TESS}, or on moderate-resolution spectroscopy (LAMOST : $\mathcal{R}\sim1,800$ and APOGEE : $\mathcal{R}\sim22,500$) that clearly preserve information correlated with these seismic quantities.}
   {With Gaia XP spectra ($\mathcal{R}\sim15$–$85$), the possibility arises to extend asteroseismic measurements to orders of magnitude more stars, despite the much lower spectral resolution. Our goal is to assess whether XP spectra retain enough information that enable reliable seismic inference for red giants.}
   {We develop hybrid Convolutional Neural Network (CNN)-Long Short-Term Memory (LSTM) models trained on red giants with seismic parameters measured from \textit{Kepler} photometry. The networks learn the subtle spectral signatures — imprinted through global stellar properties — that correlate with $\Delta\nu$, $\nu_{\text{max}}$, and $\Delta\Pi_{1}$.}
   {The models recover all three global asteroseismic parameters from Gaia XP spectra with accuracies comparable to results based on moderate-resolution surveys such as LAMOST, demonstrating that even low-resolution spectrophotometry carries sufficient information for seismic prediction. Saliency analysis reveals wavelength regions most strongly associated with seismic sensitivity and highlights physically distinct spectral behaviour between RGB and RC stars. Applying our models to Gaia DR3 yields seismic predictions for more than 2.5 million bright red giants, enabling population-level asteroseismic studies on an unprecedented scale. We also identify a small subset of low-$\Delta\nu$ red clump candidates showing unusual spectral-seismic correlations, offering new avenues for investigating evolved stellar populations.}
   {}


\maketitle
\titlerunning{Barman et. al.}
%

\section{Introduction}

Asteroseismology serves as a powerful method to investigate the internal structure of stars by analyzing their oscillations. These oscillation patterns are extracted from the light curves of stars, collected by telescopes such as CoRoT \citep{2006ESASP1306...33B}, Kepler \citep{2004ESASP.538..177B,2010Sci...327..977B}, TESS \citep{2015JATIS...1a4003R}, among others. In recent decades, numerous studies have focused on extracting valuable insights \citep{2011Natur.471..608B,2012Natur.481...55B,2014A&A...572L...5M,2019ARA&A..57...35A, 2022Natur.610...43L} from these oscillation spectra .

Primarily, stars exhibit two types of oscillation modes, $p$-modes and $g$-modes. The $g$-modes, which have buoyancy as their restoring force, are typically confined to the radiative zones of stars. The $p$-modes on the other hand have pressure as their restoring force and are confined to the stellar convective zones. In more evolved stars such as red giants, distinctive oscillations known as \textit{mixed} modes arise due to the interaction between $p$ and $g$ modes \citep{1989nos..book.....U}. Consequently, red giants offer a unique opportunity to explore stellar interiors through the $g$-mode characteristics of mixed modes. 
Crucial parameters derived from red-giant oscillation spectra include the large frequency separation ($\Delta \nu$), frequency at maximum power ($\nu_{\text{max}}$), and asymptotic period separation in mixed modes ($\Delta \Pi _l$). 

These parameters exhibit correlations with various stellar properties. For instance, $\Delta \nu$ is directly proportional to mean stellar density ($\langle\rho\rangle$), $\nu_{\text{max}}$ and scales as $\, gT_{\text{eff}}^{-\frac{1}{2}}$ \citep[where $g$ and $T_{\rm{eff}}$ are the surface gravity and effective temperature respectively;][]{2011A&A...530A.142B}, and $\Delta \Pi_{1}$ is indicative of the properties of the stellar radiative zone \citep{2013ApJ...766..118M}.

Notably, red giants exhibit two distinct phases: the red giant branch (RGB) phase when hydrogen burns in a shell around the helium core, and the red clump (RC) phase where He undergoes fusion in the core following the He flash. Despite their similar surface properties, RGBs and RCs may be differentiated based on their $\Delta \Pi _1$ values, where RGBs have $\Delta \Pi_1 < 150$ s and RCs have $\Delta \Pi_1 > 150$ s \citep{2011Natur.471..608B}. Therefore, precise measurements of these asteroseismic parameters are crucial for deducing various related stellar parameters to determine the evolutionary stages.

Numerous methodologies have been devised to accurately ascertain these asteroseismic parameters from power spectra \citep{ 2016A&A...588A..87V, Yu_2018, dhanpal2022measuring, 2024MNRAS.530.3477S}. 
However, the precise determination of these parameters requires high-quality, years-long light-curve data, imposing limitations on the number of observable red giants within a specific timeframe. Other constraints such as signal-to-noise of the oscillation spectra and stellar brightness further reduce the population size amenable to these analyses.

Various studies have indicated that photospheric abundances ought to mirror the stellar interior due to the impact of additional mixing on the upper red-giant branch \citep{2008AJ....136.2522M, 2015MNRAS.453.1855M, 2017MNRAS.464.3021M, 2017A&A...597L...3M, 2018ApJ...858L...7T}. Consequently, deriving asteroseismic parameters directly from stellar spectroscopy becomes very important.

APOGEE ($\mathcal{R} \sim 22,500$) and LAMOST ($\mathcal{R} \sim 1800$) spectra have been used to predict these asteroseismic parameters, revealing that various spectral abundances provide valuable information beyond fundamental stellar parameters \citep{2018ApJ...853...20H,2018ApJ...858L...7T, wang2023precise}.

Here, we solely use Gaia XP spectra as the input observations, characterized by coarse resolution (approximately 20 times lower than LAMOST) and unresolved features. We demonstrate that, in terms of precision regarding seismic predictions, the predicted asteroseismic parameters using Gaia XP spectra can yield comparable results to those predicted using LAMOST spectra. Furthermore, Gaia DR3 contains $\sim$ 17 million red giants \citep{Andraeetal2023a} whereas LAMOST only has $\sim$ 1 million \citep{wang2023precise}. This enables us to deduce asteroseismic parameters for a larger sample of red giants, especially with Gaia Data Release 4 (DR4), slated for mid 2026, which will have XP spectra for an order of magnitude more sources than DR3 and also cover fainter objects \citep{2024A&A...684A..29V}. Additionally, our work will also provide a large sample of RC stars which can be used to reveal the three-dimensional structure of our galaxy.

\section{Data}

Gaia DR3 provides approximately 220 million low-resolution spectra with a resolving power of $\mathcal{R} \sim 15$-$85$ \citep{Andraeetal2023b}. These spectra were obtained using two photometers: the Blue Photometer (BP), covering 330-680~nm, and the Red Photometer (RP), covering 640-1050~nm. Together, they span a wavelength range of approximately 330-1050~nm, yielding the so-called BP/RP or XP spectra. The XP spectra derived from Gaia have attracted considerable interest for stellar parameter inference, as originally explored by \citet{2012MNRAS.426.2463L}. 

Using a forward Bayesian modeling framework, \citet{Andraeetal2023b} developed the Gaia General Stellar Parameterizer from Photometry (\texttt{GSP-phot}), which provides a homogeneous catalog of effective temperatures ($T_{\rm eff}$), surface gravities ($\log g$), and overall metallicities for stars with $G < 19$. Despite their low spectral resolution, Gaia XP spectra have also been shown to retain sensitivity to chemical abundance information. While detailed elemental abundance patterns cannot be robustly recovered, numerous studies employing machine-learning and neural-network–based approaches have demonstrated that atmospheric parameters, global metallicities, $\alpha$-element abundances, and carbon abundances can be inferred from XP spectra with useful precision \citep{Andraeetal2023a,2024ApJS..272....2L,2024A&A...691A..98K,2024arXiv241016081B,2025ApJ...980...90H,2025ApJS..279....7Y,2025MNRAS.537.1984A}.

The precision of these inferred abundance estimates is not fundamentally limited by the spectral resolution, provided that exposure times and pixel-level detections are consistent. In addition, the high signal-to-noise ratio per pixel and the broad wavelength coverage of Gaia XP spectra further enable robust estimation of integrated chemical properties.

\subsection{Sample Selection}
\label{sec:SampleSelection}

\begin{figure*}
    \centering
    \begin{subfigure}[t]{0.44\textwidth}
        \centering
        \includegraphics[width=\textwidth]{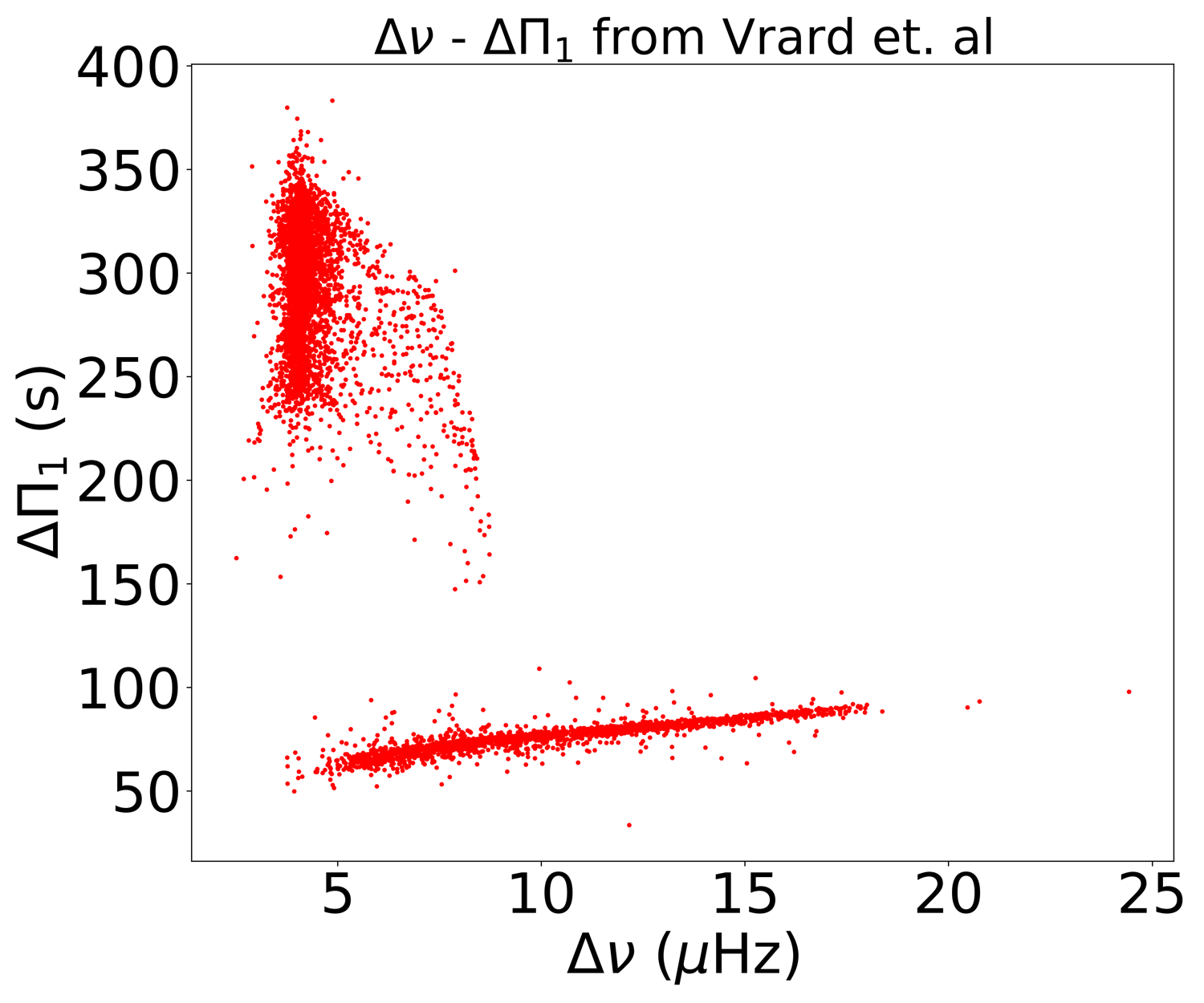}
        \caption{The distribution of $\Delta \Pi_{1}$ and $\Delta \nu$ from \citet{Vrard_2016}. We have only used the $\Delta \Pi_{1}$ values from this distribution to train our sample.}
    \end{subfigure}
    \hfill
    \begin{subfigure}[t]{0.44\textwidth}
        \centering
        \includegraphics[width=\textwidth]{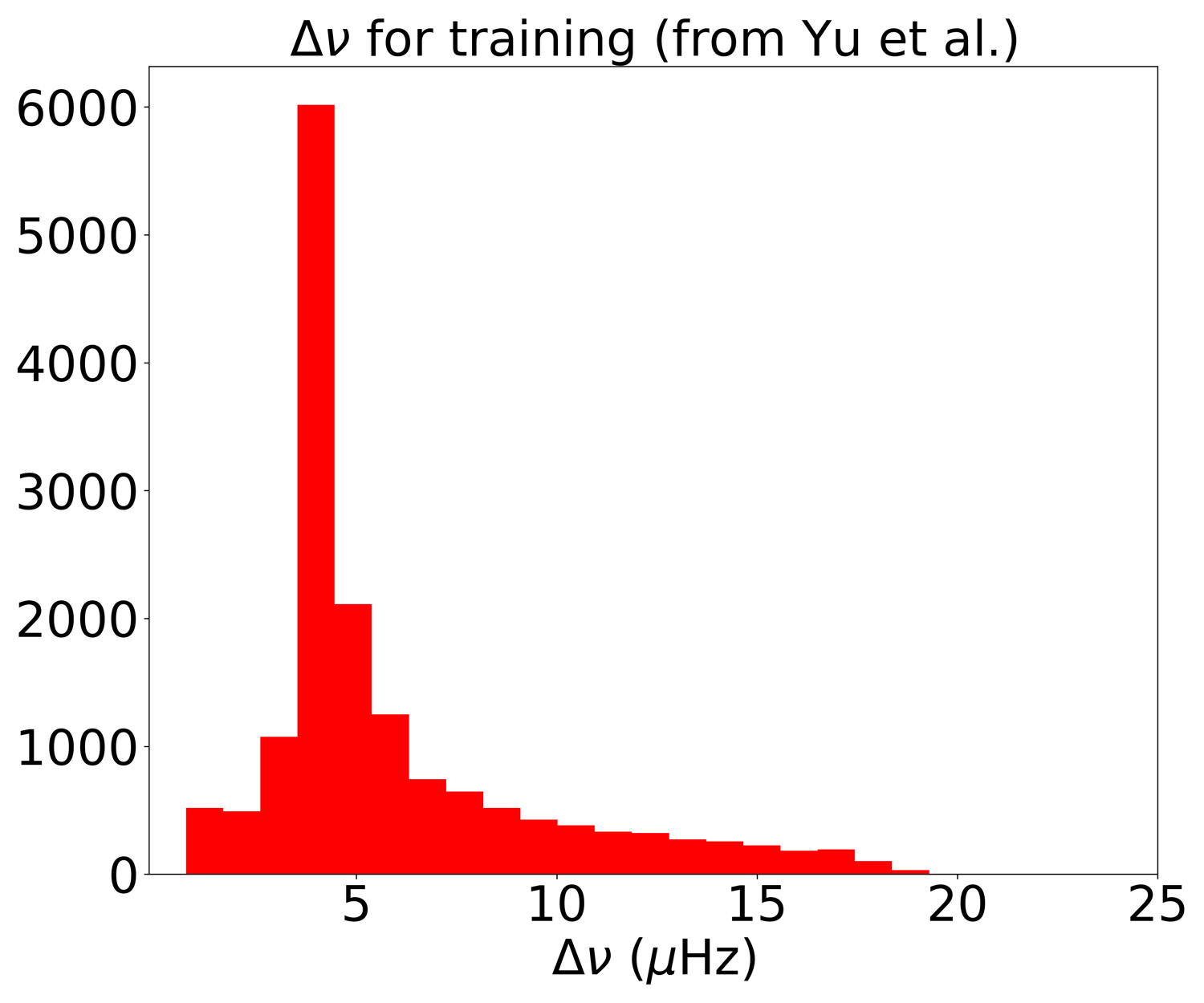}
        \caption{The histogram for the values of $\Delta \nu$ in our training data. These values are taken from \citet{Yu_2018}.}
    \end{subfigure}
    \caption{Asteroseismic parameters used in our training set.}
    \label{fig:dnu_dpi_vrard}
\end{figure*}

We used three external catalogs containing red giant data from Kepler and Gaia DR3 for this study. Specifically, we obtained measurements of $\Delta \nu$ and $\nu_{\text{max}}$ from the work of \citet{Yu_2018}. To acquire $\Delta \Pi_1$ values, we utilized the catalog by \citet{Vrard_2016}. 
The distribution  of the training data has been shown in Figure \ref{fig:dnu_dpi_vrard}.
For cross-matching the corresponding Gaia DR3 IDs for Kepler red giants, we selected only stars that had a single counterpart within a 1'' radius of the given KICs. High-confidence matches were ensured by verifying that the angular separation between Gaia and Kepler was less than 1 arcsecond and that the $(G - K)$ color was less than 2 mag. Applying these criteria, we successfully identified Gaia counterparts for 16,053 of the 16,094 Kepler red giants listed in \citet{Yu_2018}. Using the Gaia DR3 identifiers, we employed the \href{https://gaia-dpci.github.io/GaiaXPy-website/}{\texttt{GaiaXPy}} package \citep{Montegriffo_2023} to retrieve calibrated spectra for each red giant, covering wavelengths from 336 nm to 1021 nm at 2 nm resolution. We focused on bright red giants ($G$ band magnitude $< 16$) for the purpose of predictive modeling. We also ensured that our training sample has high-quality data by selecting stars with \texttt{phot\_bp\_mean\_flux\_over\_error} $>$ 10 and \texttt{phot\_rp\_mean\_flux\_over\_error} $>$ 10. We have also corrected systematic biases in the Gaia XP spectra by applying the methods described in \citet{2024ApJS..271...13H}, which are valid in the range - 0.5 $<$ $BP - RP$ $<$ 2, 3 $<$ $G$ $<$ 17.5, and $E(B - V)$ $<$ 0.8 \citep{1998ApJ...500..525S}.  

We found that the distribution of stars common to both Kepler and Gaia becomes very sparse beyond the reddening values $E(G_{BP} - G_{RP})\sim$ 0.35, constituting only a small fraction (5\%-6\%) of the total dataset. Stars with higher $E(G_{BP} - G_{RP})$ are distributed nearly uniformly over the $\Delta \nu$ and $\Delta \Pi_1$ ranges. As a result, they do not introduce systematic biases during training. While higher reddening may introduce some level of noise, the machine-learning model appears capable of accounting for this and isolating the patterns of interest to us.
Additionally, we compared our results with seismic inferences and LAMOST spectroscopic predictions from \citet{dhanpal2022measuring} ($\Delta\nu_{p}$ and  $\Delta\Pi_{p}$) and \citet{wang2023precise} ($\Delta\nu_{\rm{LAMOST}}$ and $\Delta\Pi_{\rm{LAMOST}}$), respectively.

\section{Methods}

Deep learning has emerged as a tool with which to discern general patterns from vast datasets, a capability particularly advantageous in astronomy due to the sheer volume of available data \citep{charbonneau1995genetic,ivezic2014statistics}. Depending on the specific problem at hand, deep-learning methodologies can be harnessed for either pattern classification or numerical prediction tasks. Our study employs two specialized deep-learning architectures: CNNs \citep{lecun1995convolutional} and LSTMs \citep{hochreiter1997long} networks. CNNs excel at extracting spatial features from input data, while LSTM networks are adept at capturing long-term dependencies within sequential data. The combination of these two architectures forms a potent hybrid network, particularly suited for analyzing sequential data such as the Gaia XP spectra.

\begin{figure*}[t]
    \centering
    \includegraphics[width=0.8\textwidth]{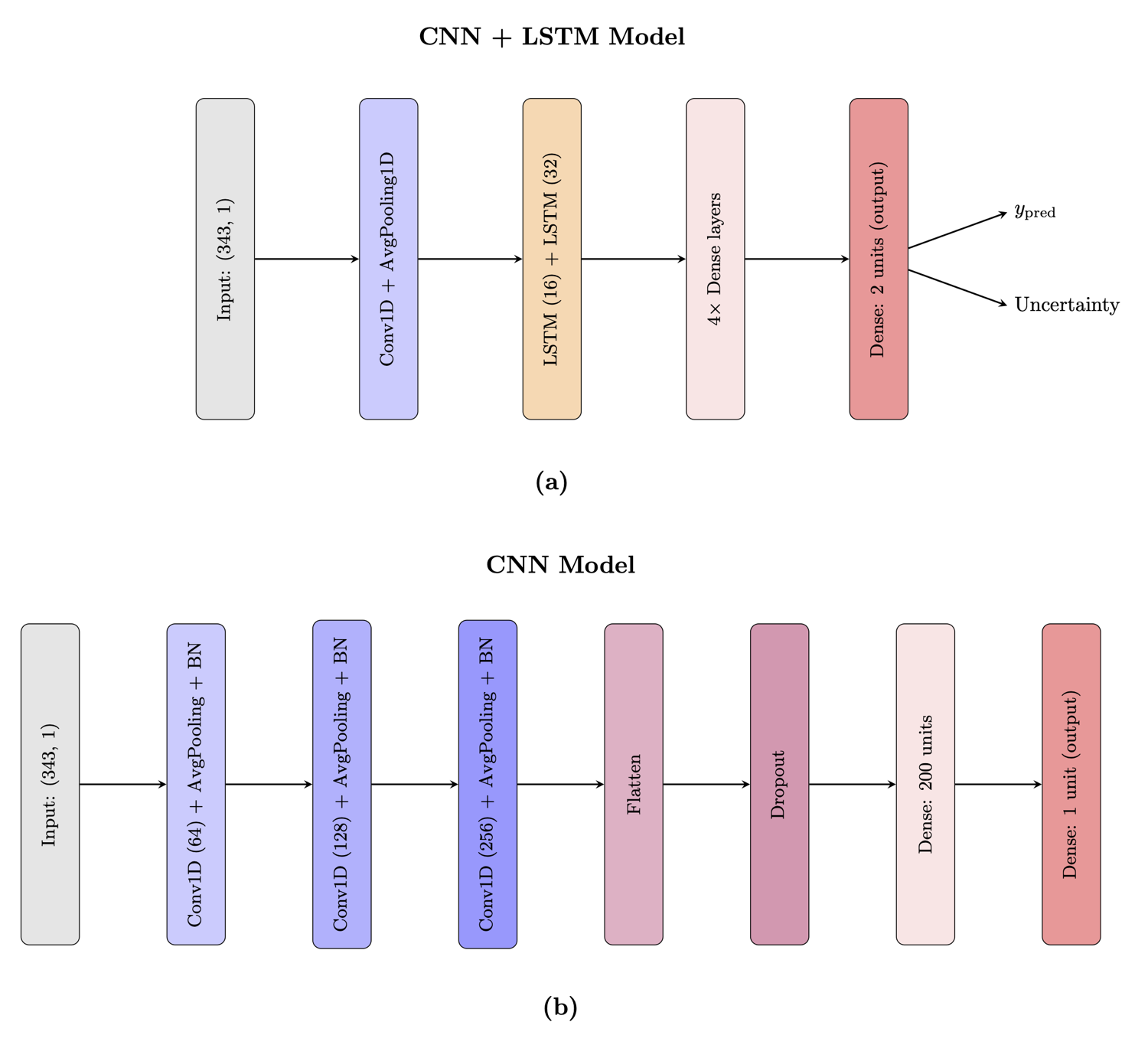}
    \caption{\textit{Model (a)} :Neural-network architecture used to output $\Delta \nu$ values. The model takes normalized Gaia XP spectra as input and computes $\Delta \nu$ and the associated error as defined in the loss function. The \texttt{CNN} layers, \texttt{LSTM} units, and the \texttt{Dense} layers are arranged in sequence as shown. A distinct, independent model with the same architecture is used to predict $\nu_{\rm{max}}$. \textit{Model (b)}: Model used to infer $\Delta \Pi_1$ from normalized Gaia XP spectra. While LSTM layers were tested, the optimal configuration uses three \texttt{CNN} blocks. The \texttt{Dropout} layer enables the use of \textit{Monte Carlo Dropout}, providing an estimate of prediction uncertainty.}
    \label{fig:model_description}
\end{figure*}

\subsection{Model Selection and Description}

The choice of model architecture in deep learning is inherently heuristic and is guided primarily by empirical performance rather than by a unique theoretical prescription. Because deep neural networks optimize highly non-convex objective functions, multiple distinct architectures can achieve comparably good solutions, and the practical aim is therefore not to identify a provable global optimum but to obtain a model that generalizes reliably to unseen data \citep{DBLP:journals/corr/abs-1206-5533, Goodfellow-et-al-2016}. In this spirit, our objective was not to determine an absolutely optimal architecture, but to select a configuration that is demonstrably sufficient for accurate and robust inference of the asteroseismic parameters. To this end, we systematically explored a range of candidate architectures and hyperparameters, and adopted the model that consistently yielded low validation loss and strong performance across our evaluation metrics while exhibiting minimal overfitting (see Appendix).

In Figure \ref{fig:model_description}, we present the architectures employed for predicting for the asteroseismic parameters discussed. We train separate models for $\Delta \nu$ and $\Delta \Pi_{1}$ in order to utilize the largest available training samples for each parameter. The $\Delta \nu$ measurements are taken from \citet{Yu_2018} ($\sim$16,000 stars), while $\Delta \Pi_{1}$ values are drawn from \citet{Vrard_2016} ($\sim$6,000 stars). Although a substantial number of stars have measurements of both quantities, adopting a joint model would require restricting the training to only those stars with simultaneous labels, thereby reducing the effective training set for each parameter relative to what is otherwise available. We note that $\Delta \nu$ and $\Delta \Pi_{1}$ exhibit a correlation in observational data and can, in principle, be modeled jointly. Such an approach is particularly more useful in data-limited regimes, where shared representations may improve performance. For example, \citet{wang2023precise} train a single model on a LAMOST spectra sample of just $\sim$ 1,800 stars with both parameters measured. In the present case, however, we prioritize maximizing the use of all available measurements for each parameter and therefore adopt separate models. These models are designed to operate on the Gaia XP spectra, which is normalized by its peak flux, as input.  
For NN model employed for estimating $\nu_{\rm{max}}$ and $\Delta\nu$ , initial layer consists of a one-dimensional CNN with 256 filters, utilizing a kernel size of 20 and a rectified linear unit (ReLU) activation function. This configuration enables the extraction of local patterns and correlations within the data. Subsequently, an average pooling \footnote{average pooling is preferred because it maintains spectral line shapes, whereas max pooling retains only the strongest (often noisy) pixel.} layer with a size of 3 and a stride of 3 is applied for downsampling. Following the feature extraction stage, two LSTM layers with 16 and 32 units, respectively, are employed to capture potential long-range dependencies within the input sequences. Subsequent to this, four dense layers, each employing ‘relu’ activation \citep{nair2010rectified}, facilitate hierarchical abstraction of the data. These dense layers contain the majority of the model parameters. The final layer consists of two neurons: one predicts the target variable, while the other estimates the associated uncertainty, the interpretation of which depends on the chosen loss function. 
All the hyper-parameters were selected using 5-fold cross-validation (refer Appendix \ref{app:5foldXvalidation}) on the training data (80\%) across different model architectures. We experimented with different kernel sizes (10, 20, 40) and filter configurations for the CNN models, and varied the number of units per layer (16, 32, 64) for the LSTM-included models. The best-performing configuration was selected based on the average validation loss across the 5-fold cross-validation splits and minimal overfitting. After model architecture and hyper-parameter selection, we retrained the model on the entire training set (80\%) and evaluated its performance on a held-out test set (20\%). For the $\Delta\Pi_1$ model, the LSTM layer was not required, as sufficient robustness was achieved using only CNN architectures, as demonstrated by the five-fold cross-validation results presented in Appendix~\ref{app:5foldXvalidation}.

\subsection{Loss Functions}

The primary aim of model training is to minimize the loss function across both training and test datasets. Despite employing the same optimizer, e.g., Adam (initial learning rate of 0.0001) \citep{kingma2017adammethodstochasticoptimization}, maintaining consistency in the choice of the loss function is critical, as different loss functions may yield disparate solutions for the same problem. In regression tasks, loss functions such as Mean Absolute Error (MAE) and Mean Squared Error (MSE) are typically applied \citep{9167435, jadon2022comprehensivesurveyregressionbased}. However, in these cases, the variance associated with each data point is different (known as \textit{heteroskedasticity}). Consequently, MAE and MSE fail to provide an accurate error estimate for each prediction.

To address this limitation, we employ the negative of a log-Laplacian as the loss function, defined as \citep{Claytor_2022, berzal2025dl101neuralnetworkoutputs}

\begin{equation}
\label{eq : laplacian_loss}
\mathcal{L}_1 = \ln(2b) + \frac{|y_{\text{true}} - y_{\text{pred}}|}{b},
\end{equation}
where \textit{b} denotes a measure of uncertainty. This loss function minimizes MAE, wherein the model exhibits bias towards the median of the output variable's range when confidence in a prediction is lacking. Formally, the Laplacian distribution's variance is $\sqrt{2}b$, although here, \textit{b} serves merely as an indicative measure of prediction confidence.

Recognizing that this loss function may not universally yield optimal solutions, we further explore the Gaussian loss function, which is the negative of a log-Gaussian
\begin{equation}
\label{eq : gassian_loss}
\mathcal{L}_2 = \frac{1}{2}\left(e^{-\log(\sigma ^2)}\right)(y_{\text{true}} - y_{\text{pred}})^2 + \frac{1}{2}\log(\sigma ^2),
\end{equation}

We use a constraint on the predicted log-variance ($\log \sigma^2 \leq 0$) to ensure numerical stability and prevent pathological variance inflation, where $\sigma^2$ denotes the predicted variance. As a result, $\sigma$ is bounded ($\sigma \leq 1$) and is interpreted as a bounded uncertainty proxy (or inverse confidence score), rather than a calibrated probabilistic error estimate. This formulation of the loss function bears resemblance to MSE, facilitating comparison \citep{Nix1994EstimatingTM, Hjorth2000BayesianTO}. Both loss functions offer distinct advantages \citep{DBLP:journals/corr/abs-2202-03870} and are investigated to discern their efficacy in addressing the specific problem at hand. 

\section{Results}
\label{sec:results}

We trained and validated our neural network models on the training and test data for $\Delta \nu$, $\Delta \Pi_{1}$ and $\nu_{\rm{max}}$ using the \texttt{Tensorflow} package. For evaluating our results, we adopt the following success metric \citep{Claytor_2022} $$\rm{accuracy}90 = \frac{1}{N}\sum_{i = 1}^{N}H(0.1 - \epsilon _i),$$ $$\rm{accuracy}80 = \frac{1}{N}\sum_{i = 1}^{N}H(0.2 - \epsilon _i),$$ where $\epsilon _i$ is the relative error of the $i^{th}$ data point and $H(x - \epsilon _i)$ is the Heaviside or step function. Hence, $\rm{accuracy}90$ and $\rm{accuracy}80$ are fractions of the dataset for which the model has predicted values within 10\% and 20\% of the true value, respectively. Additionally, we have also shown the Mean Absolute Percentage Error (MAPE) and Median Absolute Percentage Error (MedAPE) for train and test dataset of each parameters. The summary of these metrics is presented in the Table \ref{table:accuracy}.

\subsection{Prediction of $\Delta \nu$, $\Delta \Pi_{1}$ and $\nu_{\rm{max}}$}
\label{sec : Predcitions_params}

Upon selecting a subset of bright Gaia DR3 red giants using the selection criterion mentioned above, we obtained a sample comprising 15,725 stars with available $\Delta \nu$ values from Kepler observations \citep{Yu_2018}. The dataset is partitioned into training and testing subsets, with an 80\% to 20\% ratio, respectively. Accordingly, our model underwent training on 12,580 samples, with evaluation conducted on the remaining 3,145 samples. 
Extensive experimentation involving various loss functions led us to adopt the $\mathcal{L}_2$ loss function, which yielded superior performance metrics, enhancing both \textit{\rm{accuracy}90} and \textit{\rm{accuracy}80} across training and test datasets.

\begin{figure*}
    \centering
    \begin{subfigure}[t]{0.44\textwidth}
        \centering
        \includegraphics[width=\textwidth]{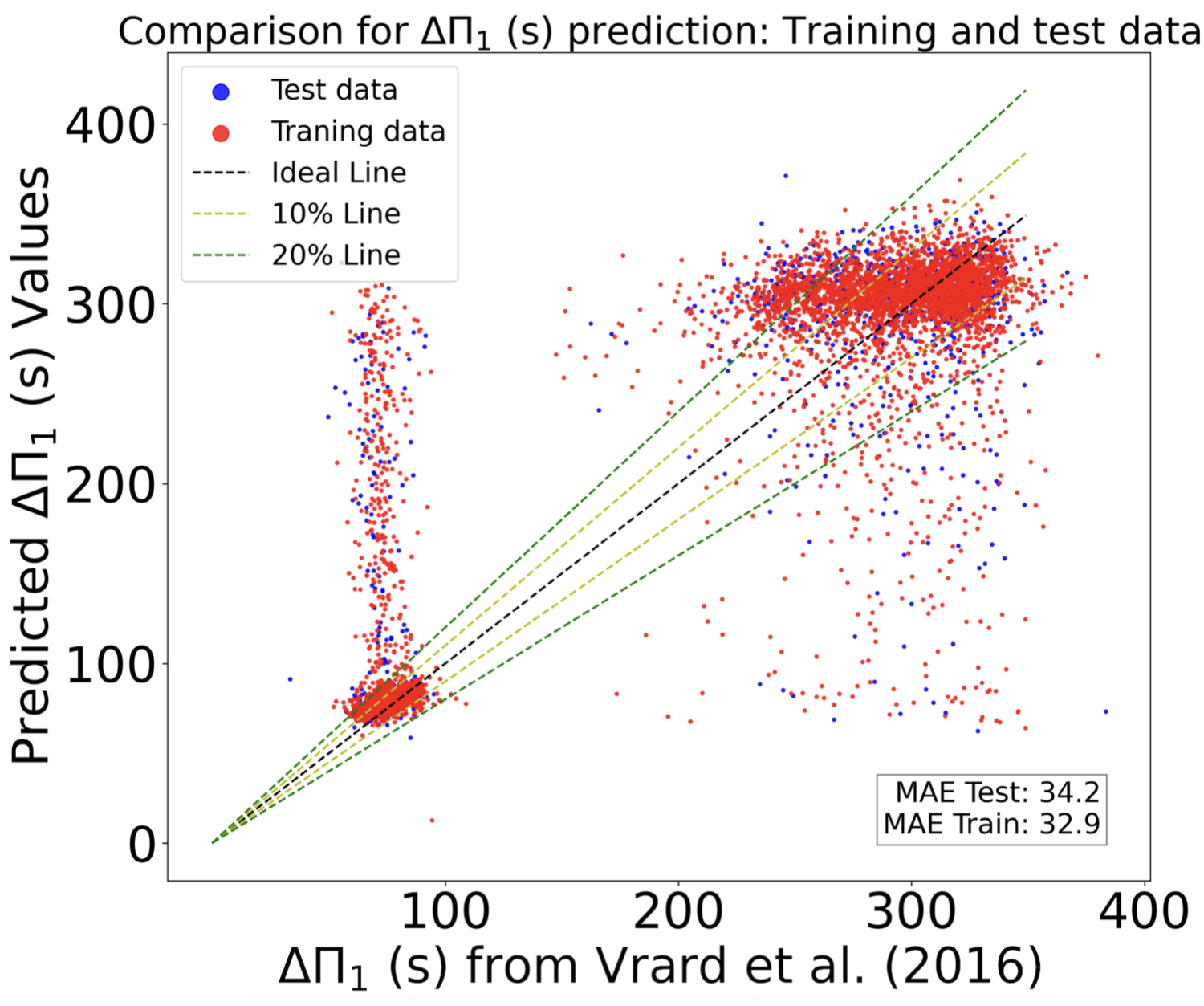}
        \caption{Measurements of $\Delta \Pi_{1}$ (in seconds) for both training and test data of Gaia red giants obtained by the neural network. The dotted red and green lines denote 10\% and 20\% error bars, respectively. This sample is taken from \citet{Vrard_2016}.}
    \end{subfigure}
    \hfill
    \begin{subfigure}[t]{0.44\textwidth}
        \centering
        \includegraphics[width=\textwidth]{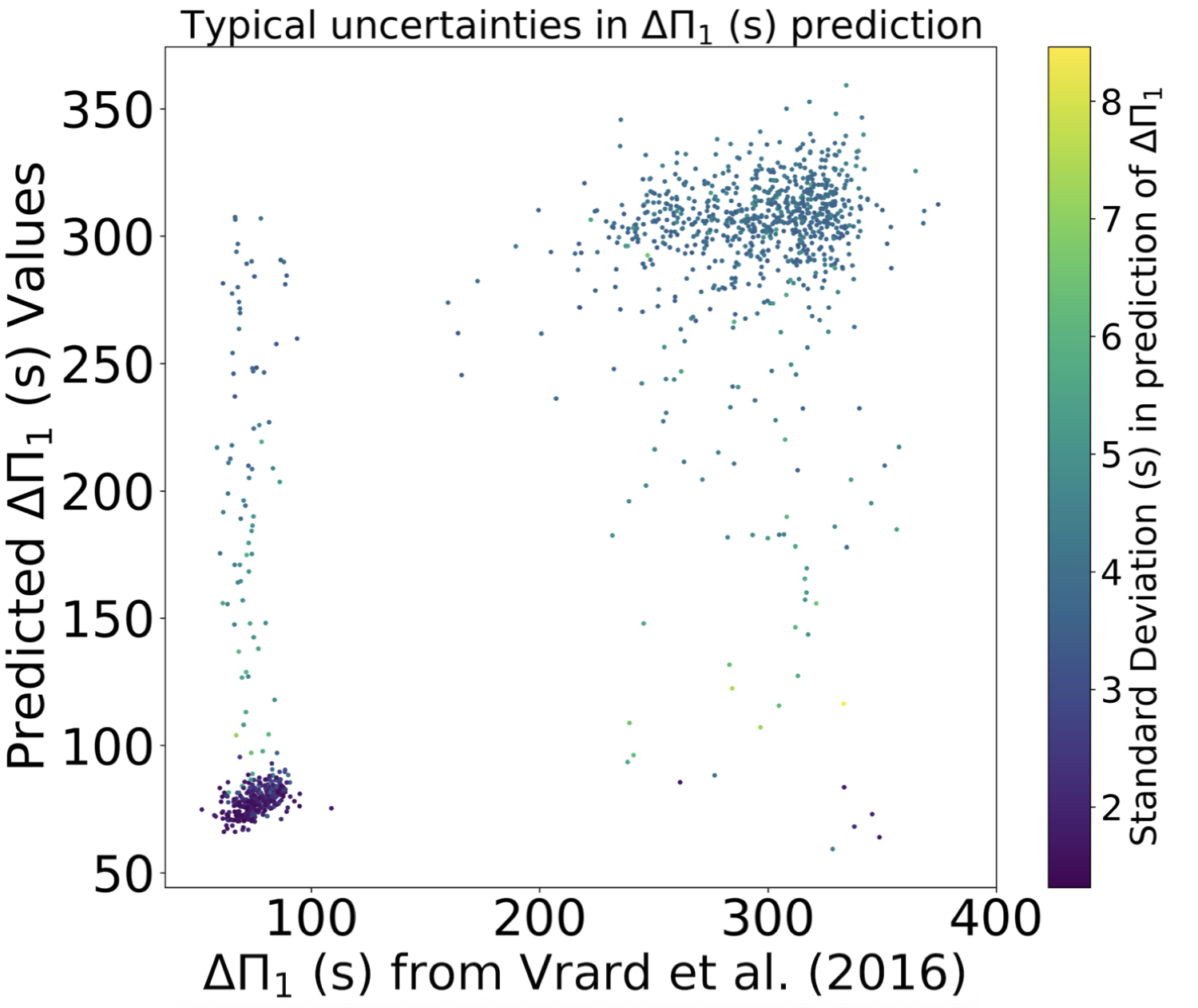}
        \caption{As our model for estimating $\Delta \Pi_{1}$ includes a \texttt{Dropout} layer, we use the Monte Carlo Dropout (MCD) technique over 100 samples to generate prediction distributions for each data point. The standard deviation associated with each distribution is shown here, offering a measure of prediction uncertainty.}
    \end{subfigure}
    \caption{Model performance on $\Delta \Pi_{1}$ prediction.}
    \label{fig:dpi_pred}
\end{figure*}

For predicting $\Delta \Pi_{1}$, our model was trained using 4,885 samples from the catalogue of \cite{Vrard_2016}, with evaluation performed on 1,221 independent test samples. Since the model incorporates a \texttt{Dropout} layer, we employed the \textit{Monte Carlo Dropout} technique \citep{gal2016dropoutbayesianapproximationrepresenting} to estimate the uncertainty in the predictions. During training, we found that employing Mean Absolute Error (MAE) as the loss function provided the most effective results (Table \ref{table:accuracy}), in contrast to the more conventional $\mathcal{L}_1$ or $\mathcal{L}_2$ losses, which yielded maximum \textit{\rm{accuracy}90} values of only 20–30\%.
Random predictions from this dataset would achieve an \textit{\rm{accuracy}90} of approximately 21\% and an \textit{\rm{accuracy}80} of approximately 40\%, highlighting the superior performance of our trained model (Figure \ref{fig:dpi_pred}). Our approach thus emphasizes optimizing model performance while maintaining a balanced confidence estimate. The training samples for both $\Delta \nu$ and $\Delta \Pi_{1}$ cover stars with astrophysical parameters within the range of:

\begin{itemize}
    \item[(a)] 3956.5 K $\le$ $T_{\rm{eff}}$ $\le$ 5181.0 K, -0.7 \texttt{dex} $\le$ $\rm{[M/H]}$ $\le$ 0.435 \texttt{dex}, 1.307 $<$ $\log g$ $<$ 3.47.
    \item[(b)] 4419.1 K $\le$ $T_{\rm {eff}}$ $\le$ 5148.9 K, -0.7 \texttt{dex} $\le$ $\rm{[M/H]}$ $\le$ 0.416 \texttt{dex}, 1.884 $\le$ $\log g$ $\le$ 3.373.
\end{itemize}

\begin{figure*}[t]
    \centering
    \includegraphics[width=0.99\textwidth]{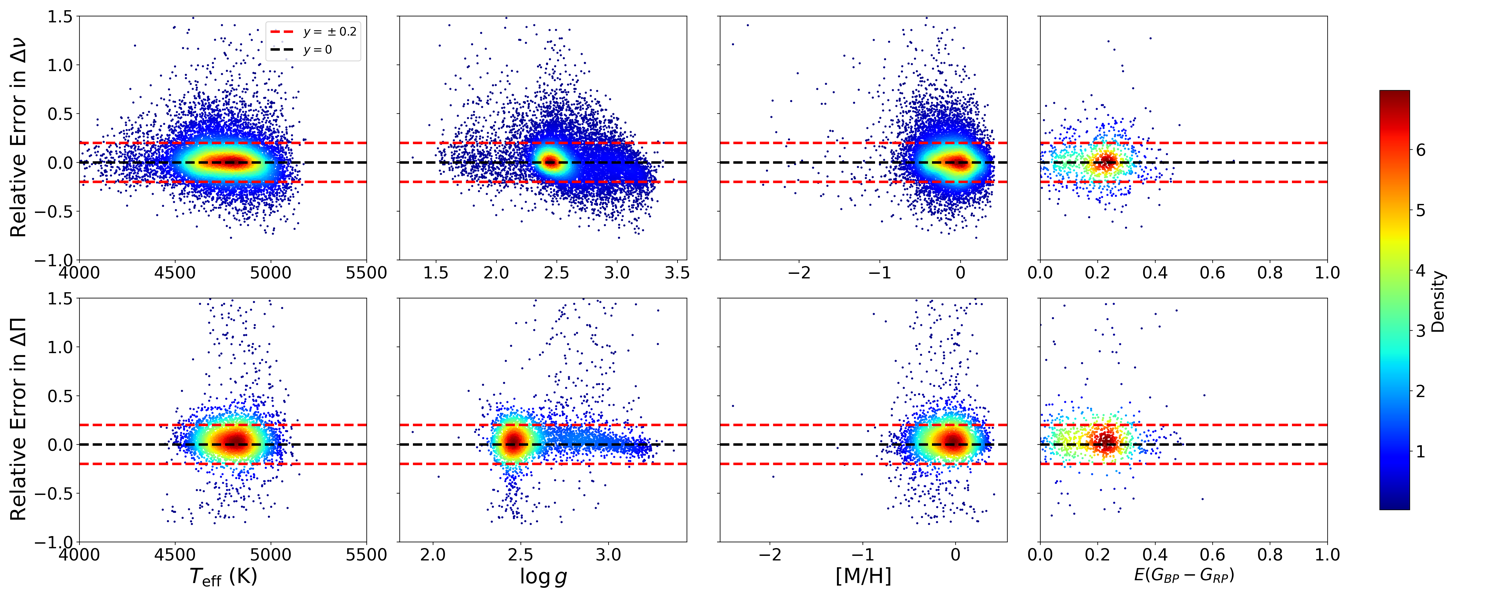}
    \caption{The first and second rows show the relative errors in the predicted $\Delta \nu$ and $\Delta \Pi_{1}$ across various astrophysical parameters — $\rm{T}_{\rm{eff}}$, $\log(g)$, [M/H], and reddening ($E(G_{BP} - G_{RP})$). The values of the astrophysical parameters are taken from \citet{Andraeetal2023a}. The last column in each row corresponds to reddening, which is obtained from the \texttt{gaia.source} table using \texttt{astroquery.gaia}. As shown, higher reddening does not always correlate with worse predictions; many $\Delta \nu$ and $\Delta \Pi_{1}$ values are within the 20\% error margin. The stars shown here are from the Kepler-Gaia red giant training sample.}
    \label{fig:error_astro_params}
\end{figure*}

The distribution of these astrophysical parameters with the associated errors in predictions from the models can be found in the Figure \ref{fig:error_astro_params}. These values of $T_{\rm{eff}}$, $\log g$ and $\rm{[M/H]}$  are taken from \citet{Andraeetal2023a}.

We draw on the power-law relationship elucidated by \cite{Stello_2009}, which establishes a functional linkage between $\Delta \nu$ and $\nu_{\rm{max}}$ for solar-like oscillations $$ \Delta \nu = (0.263 \pm 0.009)\,\mu \rm{Hz} 
\left(\frac{\nu_{\text{max}}}{\mu \rm{Hz}}\right)^{0.772 \pm 0.005}.$$
 
\begin{figure*}
    \centering
    \begin{subfigure}[t]{0.42\textwidth}
        \centering
        \includegraphics[width=\textwidth]{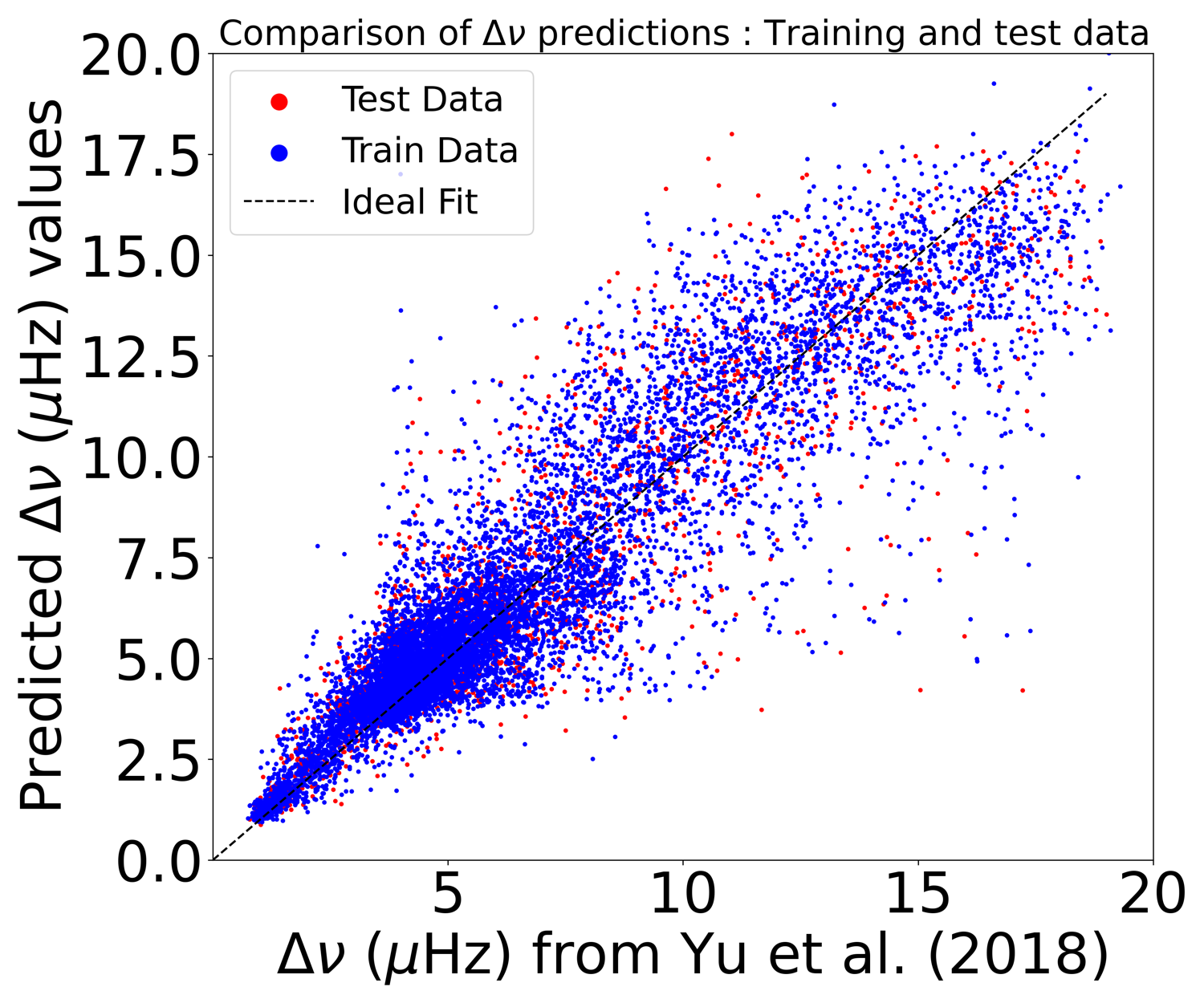}
        \caption{$\Delta \nu$ (in $\mu \rm{Hz}$) predictions from training and test data of Gaia red giants. The sample is taken from \citet{Yu_2018}.}
    \end{subfigure}
    \hfill
    \begin{subfigure}[t]{0.42\textwidth}
        \centering
        \includegraphics[width=\textwidth]{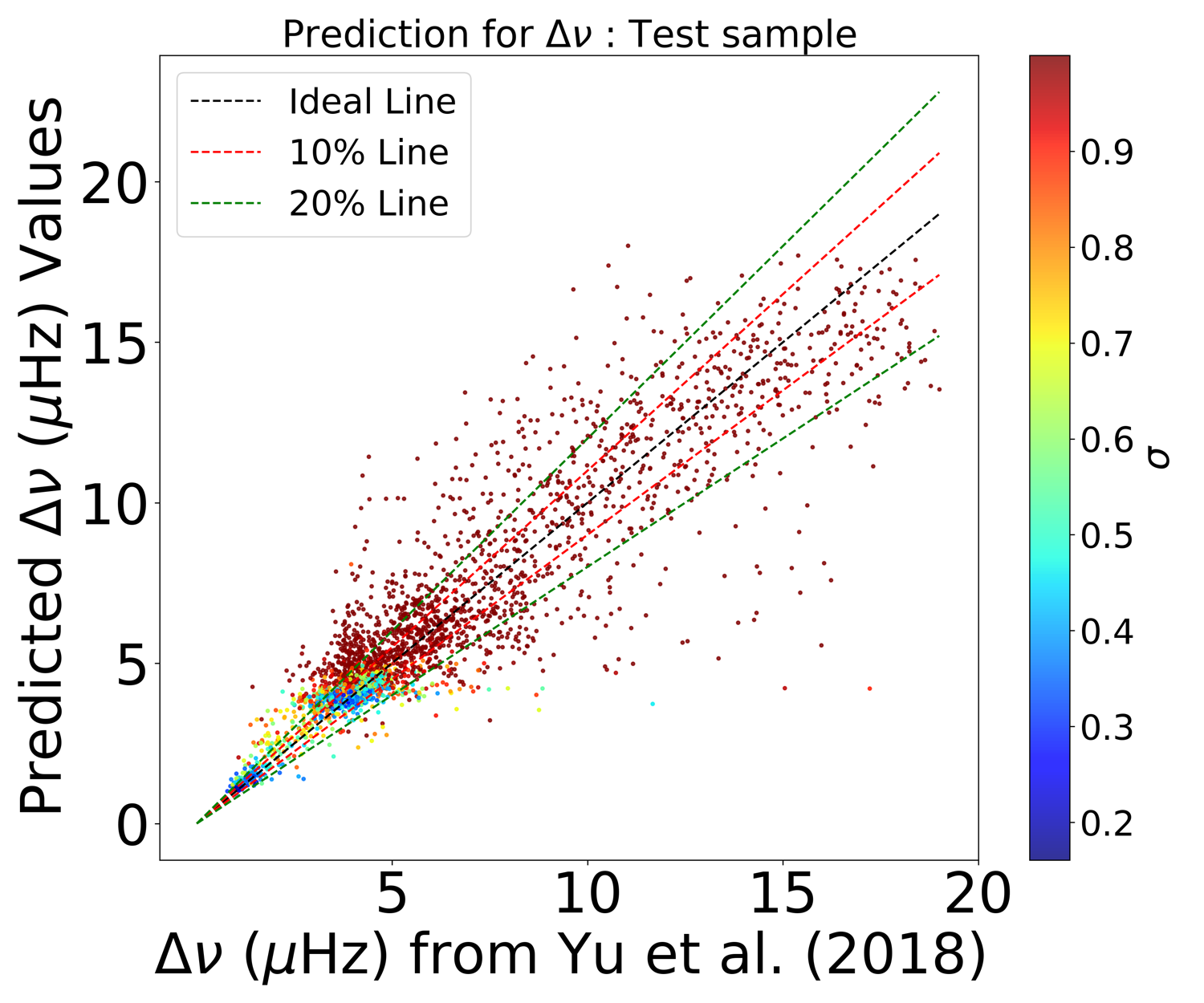}
        \caption{Predictions for test data only. Dotted red and green lines indicate 10\% and 20\% error margins, respectively. The color bar shows the standard deviation ($\sigma$) associated with each prediction.}
    \end{subfigure}
    \caption{Performance of the neural network model on predicting $\Delta \nu$ from Gaia XP spectra.}
    \label{fig:dnu_pred}
\end{figure*}

\begin{figure*}
    \centering
    \begin{subfigure}[t]{0.44\textwidth}
        \centering
        \includegraphics[width=\textwidth]{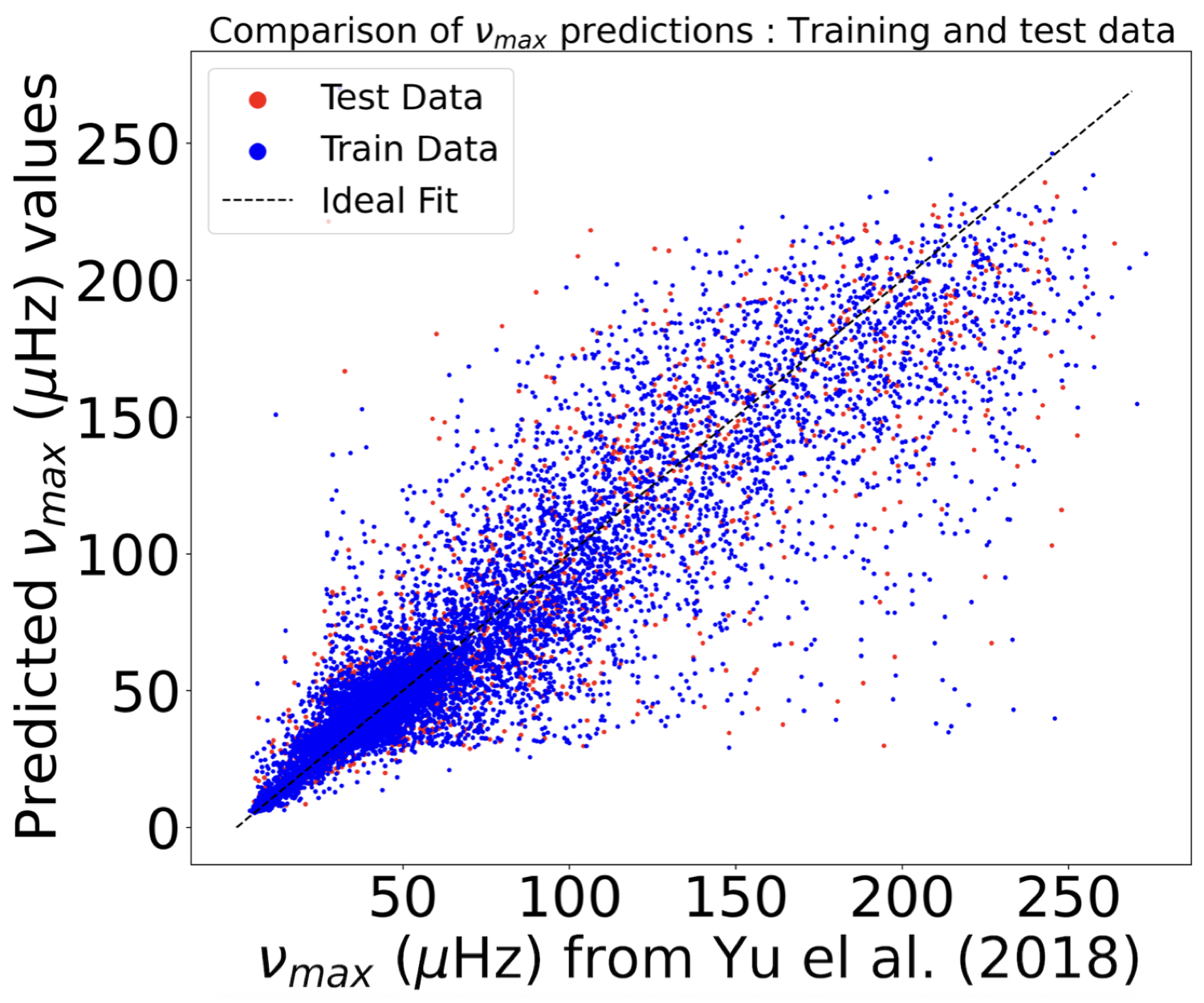}
        \caption{Predictions for $\nu_{\rm{max}}$ (in $\mu \rm{Hz}$) on both training and test data from Gaia.}
    \end{subfigure}
    \hfill
    \begin{subfigure}[t]{0.44\textwidth}
        \centering
        \includegraphics[width=\textwidth]{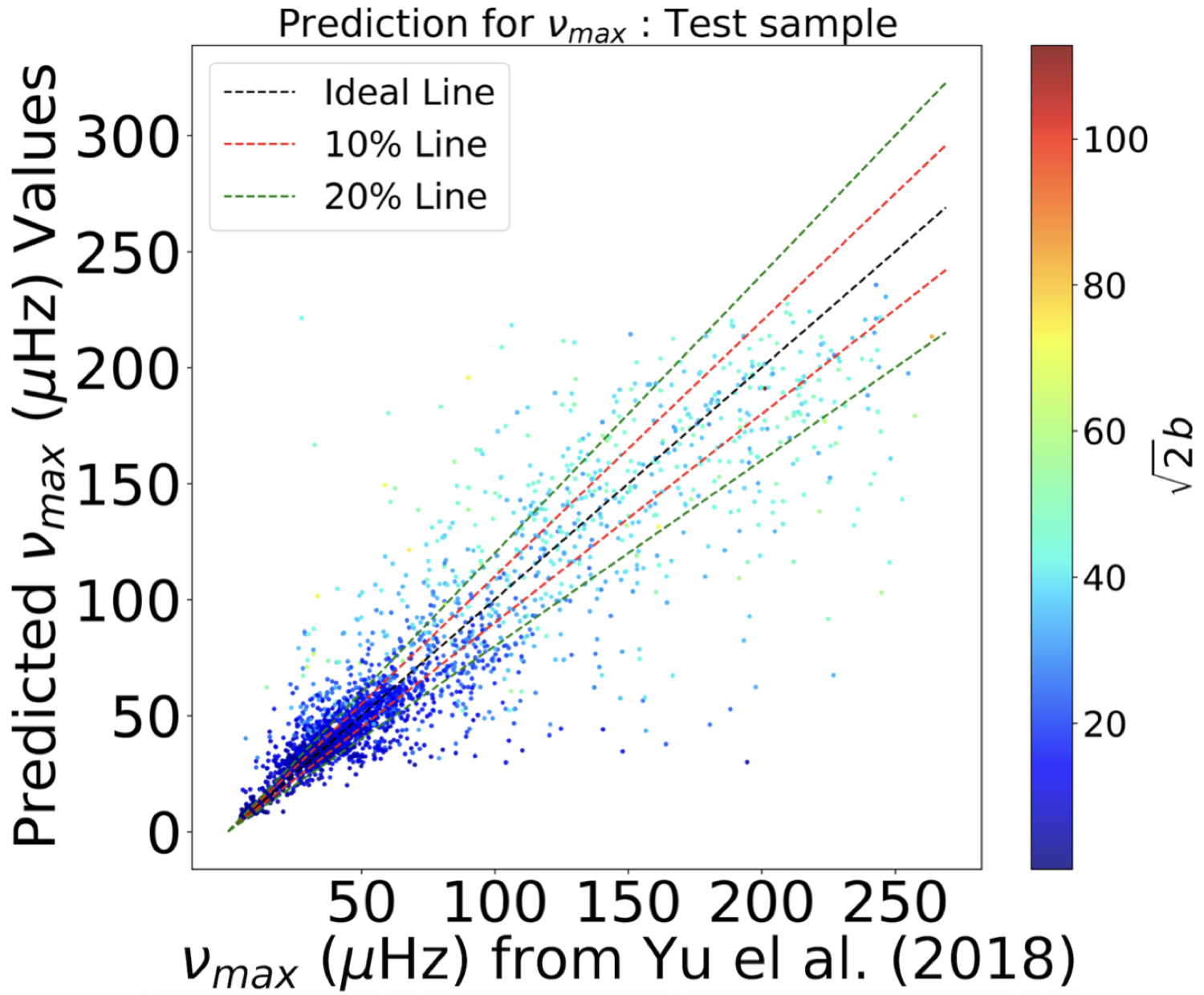}
        \caption{Predictions for test data. Dotted red and blue lines indicate 10\% and 20\% error margins, respectively. The color bar represents the associated standard deviation ($\sqrt{2}b$).}
    \end{subfigure}
    \caption{Neural network predictions of $\nu_{\rm max}$ from Gaia XP spectra.}
    \label{fig:spec_noise_plots}
\end{figure*}

\begin{figure*}
    \centering
    \begin{subfigure}[t]{0.49\textwidth}
        \centering
        \includegraphics[width=\textwidth]{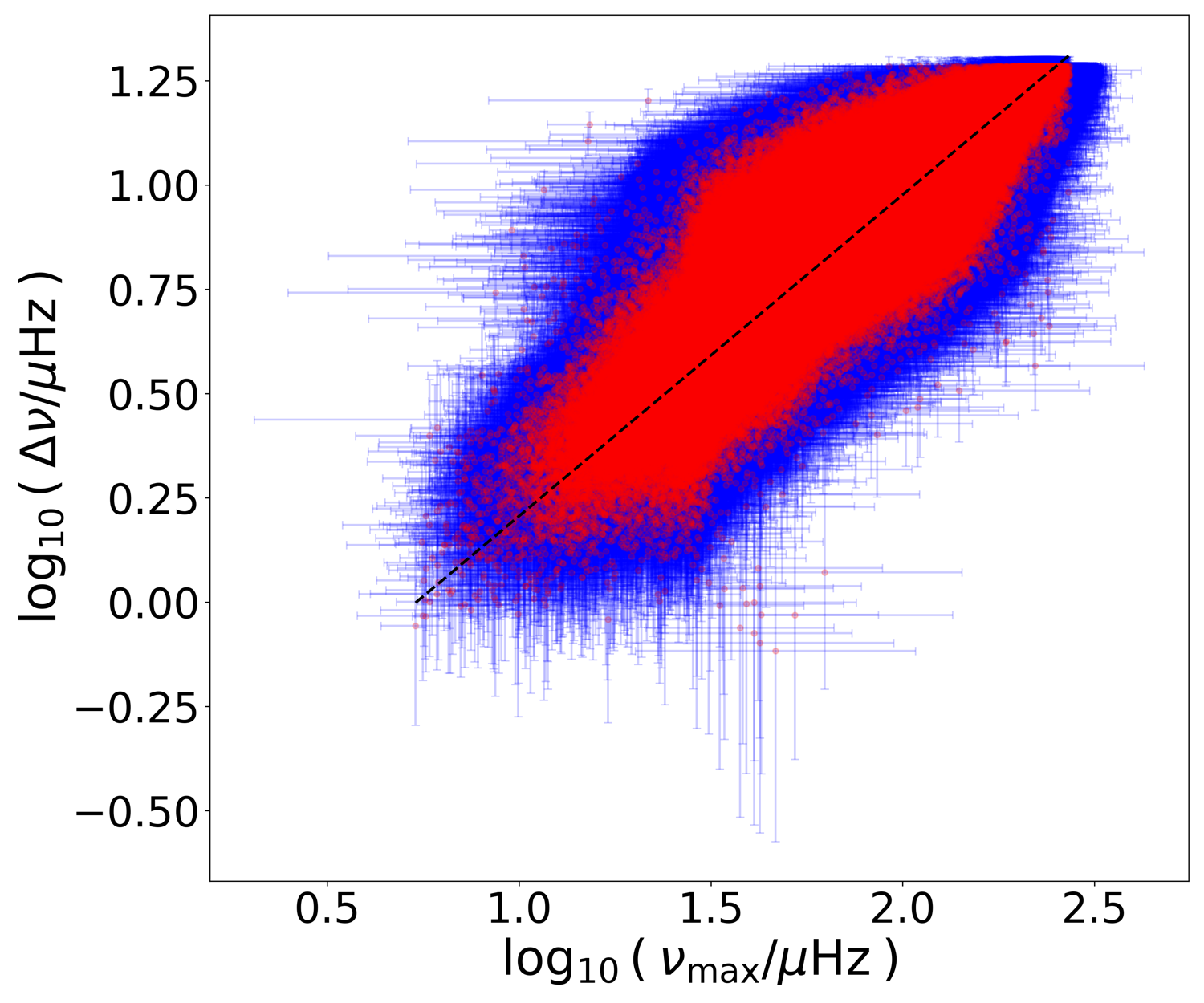}
        \caption{Predictions for all Gaia stars, as described in Section~\ref{sec: results_4.3}.}
    \end{subfigure}
    \hfill
    \begin{subfigure}[t]{0.49\textwidth}
        \centering
        \includegraphics[width=\textwidth]{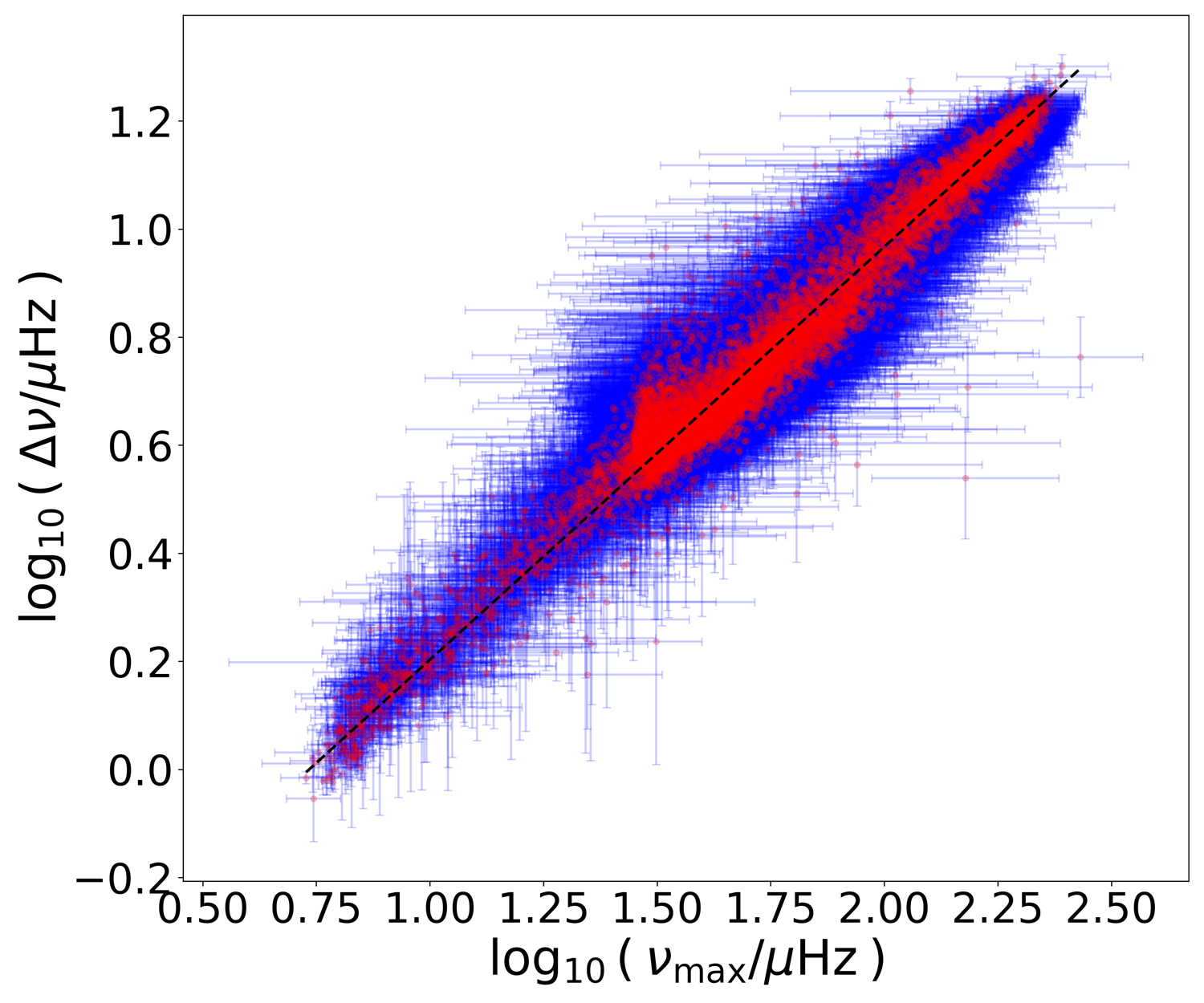}
        \caption{Predictions for stars from the \citet{Yu_2018} sample.}
    \end{subfigure}
    \caption{Plots showing $\log_{10}(\Delta \nu/\mu\rm{Hz})$ and $\log_{10}(\nu_{\rm max}/\mu\rm{Hz})$ inferred by the corresponding trained models from Gaia XP spectra.}
    \label{fig:numax_relation}
\end{figure*}

A model proficient in predicting $\Delta \nu$ with fidelity likely encodes information regarding $\nu_{\rm{max}}$ as well. Notably, $\nu_{\rm{max}}$ values tend to exhibit larger error margins than those of $\Delta\nu$. This may be due to the offset of surface gravity (which is related to 
 $\nu_{\rm{max}}$) calculated from spectroscopy and asteroseismology \citep{2017A&A...597L...3M}. Leveraging the same dataset utilized for $\Delta \nu$ predictions, we developed a dedicated model for predicting $\nu_{\rm{max}}$. To achieve this, we employed the $\mathcal{L}_1$ loss function and achieved corresponding \textit{\rm{accuracy}90} and \textit{\rm{accuracy}80} metrics of 44.1\% and 69\%, respectively. Our findings are comprehensively summarized in Table \ref{table:accuracy}. The results of our model on the training and test dataset for $\Delta\nu$ and $\nu_{\rm{max}}$ can be visualized from the Figure \ref{fig:dnu_pred} and \ref{fig:spec_noise_plots}, respectively. Using the inferences from the machine learning model, including the associated errors, we have derived the scaling relation between $\Delta\nu$ and $\nu_{\text{max}}$ in the form of $\Delta\nu = \beta.(\nu_{\text{max}})^{\gamma}$ for all the selected red giants in the Gaia DR3 catalog. A linear fit, performed using \textit{orthogonal distance regression} (implemented using \texttt{scipy.odr} \citep{boggs1989orthogonal} , yielded the values $\beta = 0.2729 \pm 0.000045$ and $\gamma = 0.7704 \pm 0.00008$ for Gaia red giants, and $\beta = 0.2745 \pm 0.00082$ and $\gamma = 0.7643 \pm 0.00093$ for the stars present in the training set which were obtained from \cite{Yu_2018}. The linear fits to $\log_{10} \left( \nu_{\rm{max}}/\mu\rm{Hz} \right)$-$\log_{10} \left( \Delta\nu/\mu\rm{Hz} \right)$ for both the the entire sample of Gaia red giants and training set are presented in Figure \ref{fig:numax_relation}.

\begin{table}
\centering
\scriptsize
\renewcommand{\arraystretch}{1.2} 
\begin{tabular}{|c | c c c|}
\hline
\textbf{} & \textbf{$\Delta \nu$} & \textbf{$\nu_{\rm{max}}$} & \textbf{$\Delta \Pi _1$} \\ \hline
\textbf{Loss function} & $\mathcal{L}_2$ & $\mathcal{L}_1$ & \textit{MAE} \\ \hline
\textbf{\rm{accuracy}90 (Train)} & 54.6\% & 46.4\% & 61\% \\ 
\textbf{\rm{accuracy}90 (Test)} & 54.9\% & 46.2\% & 59.2\% \\ \hline
\textbf{\rm{accuracy}80 (Train)} & 79.2\% & 70.8\% & 81\% \\ 
\textbf{\rm{accuracy}80 (Test)} & 78.9\% & 70.5\% & 79\% \\ \hline
\textbf{\rm{MAPE} (Train)} & 14.9\% & 18.2\% & 22.2\% \\ 
\textbf{\rm{MAPE} (Test)} & 15.2\% & 18.9\% & 24.5\% \\ \hline
\textbf{\rm{MedAPE} (Train)} & 9.9\% & 11.1\% & 7\% \\ 
\textbf{\rm{MedAPE} (Test)} & 9.9\% & 11.3\% & 8.4\% \\ \hline
\end{tabular}
\caption{Results for predictions.}
\label{table:accuracy}
\end{table}

\subsection{Saliency maps: important spectral features}

Our objective is not only to predict asteroseismic parameters, but also to identify which regions of the spectra inform us about the underlying stellar physics. For instance, if the model learns to distinguish red giant branch (RGB) stars from red clump (RC) stars, it is of particular interest to determine which spectral regions contribute most strongly to this classification. To this end, we employ the method of \textit{saliency maps} \citep{Simonyan2013DeepIC, Bhambra_2022, kechris2025timeseriessaliencymaps}, which provides a pixel-level attribution of model predictions and is applicable to any differentiable model. This approach enables us to interpret the model’s learned representations in terms of the input spectra.

Formally, given an input Gaia XP spectra $\mathbf{x} \in \mathbb{R}^d$ (where $d$ denotes the number of sampled wavelength bins in the Gaia XP spectra) and a scalar model output $f(\mathbf{x})$, the saliency score associated with the $i$-th input feature, $x_i$, is defined as the absolute value of the gradient of the output with respect to that feature:
\begin{equation}
    S_i = \left| \frac{\partial f(\mathbf{x})}{\partial x_i} \right|, \qquad i = 1, \ldots, d.
\end{equation}

The quantity $S_i$ thus measures the local sensitivity of the model output to perturbations in the $i$-th wavelength bin. Features with large $S_i$ values indicate spectral regions to which the model is highly responsive, and hence are likely to encode astrophysically meaningful information relevant for the prediction task. Conversely, features with small $S_i$ values contribute less to the model’s decision. This framework provides a direct, quantitative means of linking machine-learning predictions to physical spectral features.

\begin{figure*}[t]
    \centering
    \includegraphics[width=0.85\textwidth]{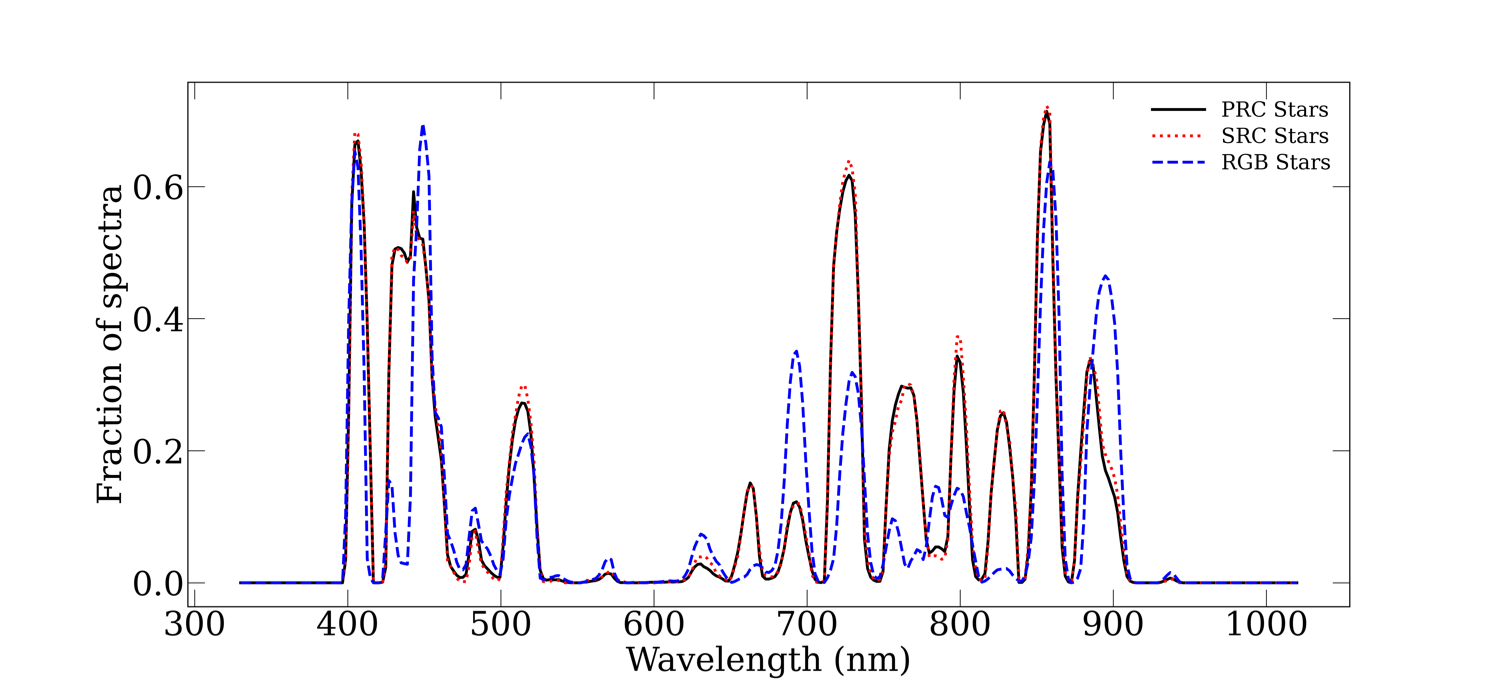}
    \caption{Saliency maps for the $\Delta\Pi_1$ model, illustrating the spectral regions most relevant for distinguishing red giant branch (RGB) stars from primary and secondary red clump stars (PRCs, SRCs). The curves show the fraction of stars whose normalized saliency scores exceed a threshold of 0.6 at each wavelength bin. While the blue portion of the spectra ($\lesssim 600$ nm) shows little distinction between evolutionary states, differences emerge in the redder regime ($600$–$850$ nm). In particular, the ranges 685–699 nm, 658–664 nm, and 820–834 nm exhibit pronounced differences: RGBs dominate in the first region, while RCs dominate in the latter two. These results indicate that the redder part of the Gaia XP spectra carries stronger discriminative power for separating RGBs from RCs.
    }
    \label{fig:saliency_dpi}
\end{figure*}

For the $\Delta\Pi_1$ model, we examined which wavelength ranges are most important in differentiating RGB from RC stars. Since the distribution of $\Delta\Pi_1$ values differs between these two evolutionary states, one would expect the model to rely on distinct spectral features when making predictions. To quantify this, we computed the fraction of RGBs and RCs whose normalized saliency scores exceeded a specified threshold (= 0.5) at each wavelength bin. The resulting distributions are shown in Figure~\ref{fig:saliency_dpi}, where the regions of high importance appear as contiguous bands across the spectral domain.

\begin{figure*}
    \centering
    \begin{subfigure}[t]{0.75\textwidth}
        \centering
        \includegraphics[width=\textwidth]{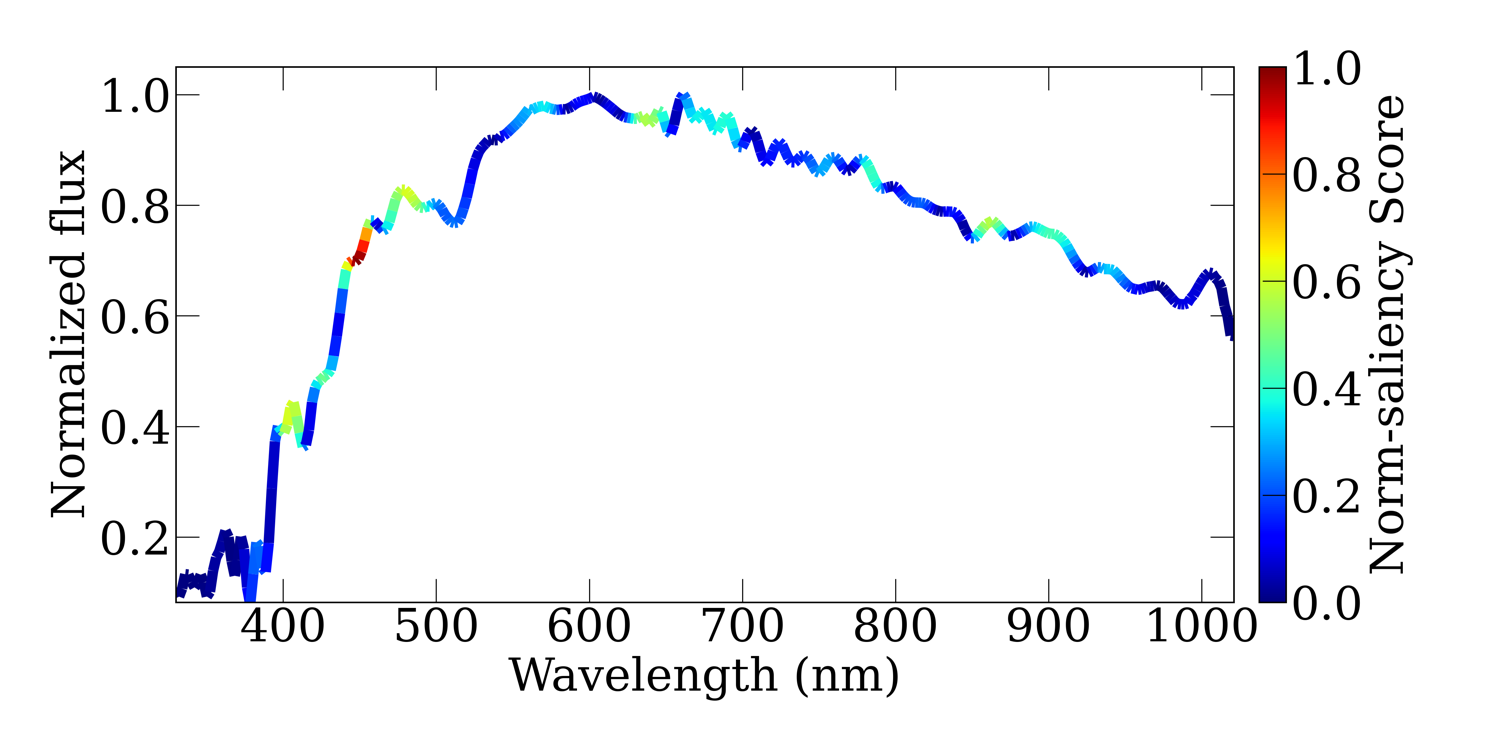}
        \caption{KIC 3215869}
    \end{subfigure}
    \vskip 0.5cm 
    \begin{subfigure}[t]{0.75\textwidth}
        \centering
        \includegraphics[width=\textwidth]{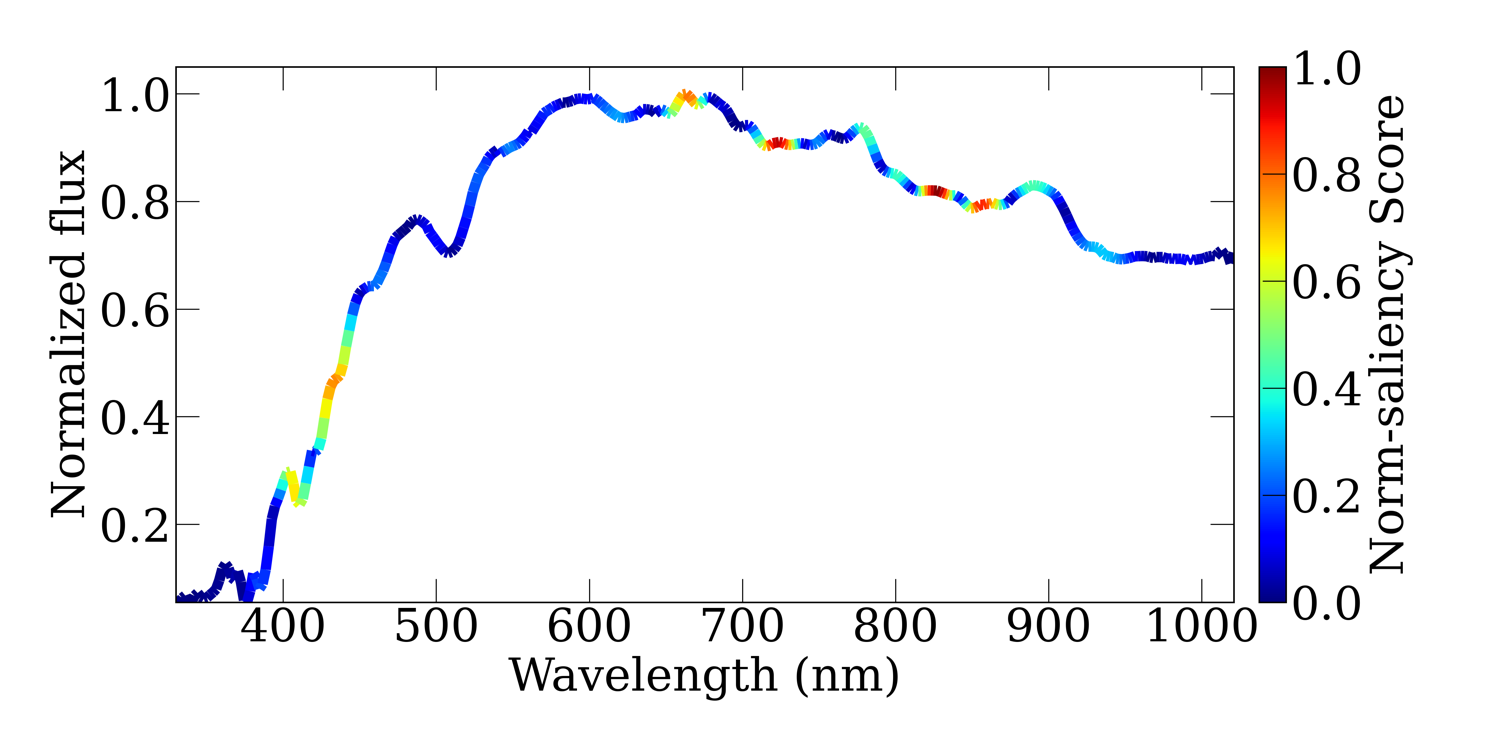}
        \caption{KIC 2016676}
    \end{subfigure}
    \caption{Normalized Gaia XP spectra color-coded with saliency scores for a representative RGB star (\textit{panel a}) and RC star (\textit{panel b}), both selected within the range $6\, \mu{\rm Hz} \leq \Delta\nu \leq 7\, \mu{\rm Hz}$. Consistent with the population-level trends shown in Figure~\ref{fig:saliency_dpi}, we see differences in the saliency distributions. In particular, the RC star exhibits systematically enhanced saliency across the redder wavelength bands, highlighting the spectral regions most influential for the model in distinguishing between evolutionary states.
    }
    \label{fig:saliency_RGB_RC}
\end{figure*}

Over the bluer portion of the spectra in Figure~\ref{fig:saliency_dpi} (up to $\sim 600$ nm), the saliency maps of RGBs, primary RCs (PRCs), and secondary RCs (SRCs) show little distinction. However, in the redder regime ($600$--$850$ nm), notable differences emerge between the saliency patterns of RGBs and RCs, while PRCs and SRCs continue to exhibit similar responses. This suggests that the model leverages different spectral regions for distinguishing RGBs from RCs (see Figure \ref{fig:saliency_RGB_RC}), whereas it produces similar responses to spectral features in the case of PRCs versus SRCs. The wavelength ranges (a) 685--699 nm, (b) 658--664 nm, and (c) 820--834 nm are particularly noteworthy. As shown in Figure~\ref{fig:saliency_dpi}, the model assigns high saliency scores in range (a) predominantly to RGB stars, with little corresponding contribution from RCs. In contrast, the opposite trend is observed in ranges (b) and (c), where RCs are emphasized far more strongly than RGBs. More generally, the saliency distribution indicates that the redder part of the spectrum ($\gtrsim 600$ nm) carries greater discriminative power for RCs, suggesting that the model relies more heavily on these wavelengths when distinguishing RCs from RGBs.

A recent high-resolution spectroscopic study by \citet{2025MNRAS.540.3919W} has shown that RGB - RC differences manifest through subtle CN/$\rm{C}_2$ band-head strengths, isotopic-ratio changes, and microturbulence-dependent line-wing variations , and abundance analyses reveal systematic C/N and Na changes across the helium-flash boundary \citep{2018ApJ...858L...7T, 2019MNRAS.482.4155L}. None of these line- or band-level signatures are preserved in Gaia XP, whose extremely low spectral resolution washes out individual CN/C$_2$ features, H-line wings, and even most blended molecular structures. Instead, XP retains only very coarse, low-frequency information: broad spectral energy distribution (SED) shape, and smooth continuum-level curvature shaped by overall opacity and temperature-gravity-composition combinations. CNN-LSTM networks are particularly effective at exploiting exactly this kind of morphology-level signal, integrating shallow, distributed flux variations across wide wavelength regions, which explains why a data-driven model can still distinguish RGB and RC stars and may naturally focus on the 5800-8450~\AA\ interval, where XP preserves the strongest large-scale flux-shape differences. In the traditional spectroscopic sense, this red-optical region may be dismissed as information-poor in some studies  \citep{2022ApJS..259...51W, wang2023precise}, but at XP resolution, these broad flux-morphology trends become precisely the features that data-driven models capitalize on.

\begin{figure*}[t]
    \centering
    \includegraphics[width=0.85\textwidth]{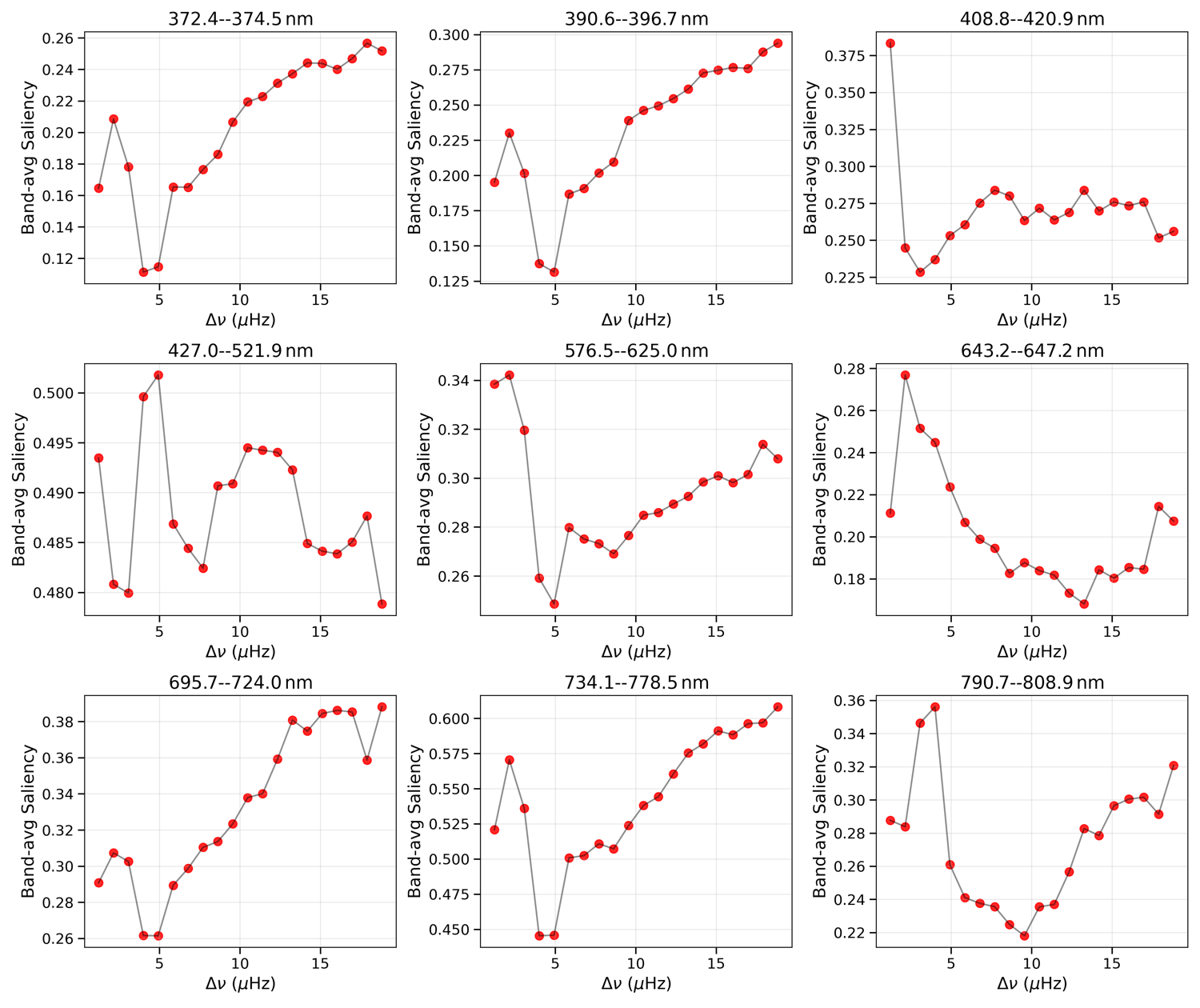}
    \caption{Average normalized saliency scores as a function of $\Delta\nu$ for selected wavelength bands in the $\Delta\nu$ model, computed using RGB stars. The $\Delta\nu$ range $0$-$20~\mu\mathrm{Hz}$ is divided into 20 bins, and the saliency scores are averaged within each bin for bands with normalized saliency values exceeding 0.1. Several wavelength intervals exhibit systematic trends with $\Delta\nu$, including a steady increase in saliency at shorter wavelengths (e.g., $372.4$-$374.5~\mathrm{nm}$), indicating differential sensitivity of specific spectral regions to stellar evolutionary state.}
    \label{fig:saliency_dnu}
\end{figure*}

\begin{figure*}[t]
    \centering
    \includegraphics[width=0.85\textwidth]{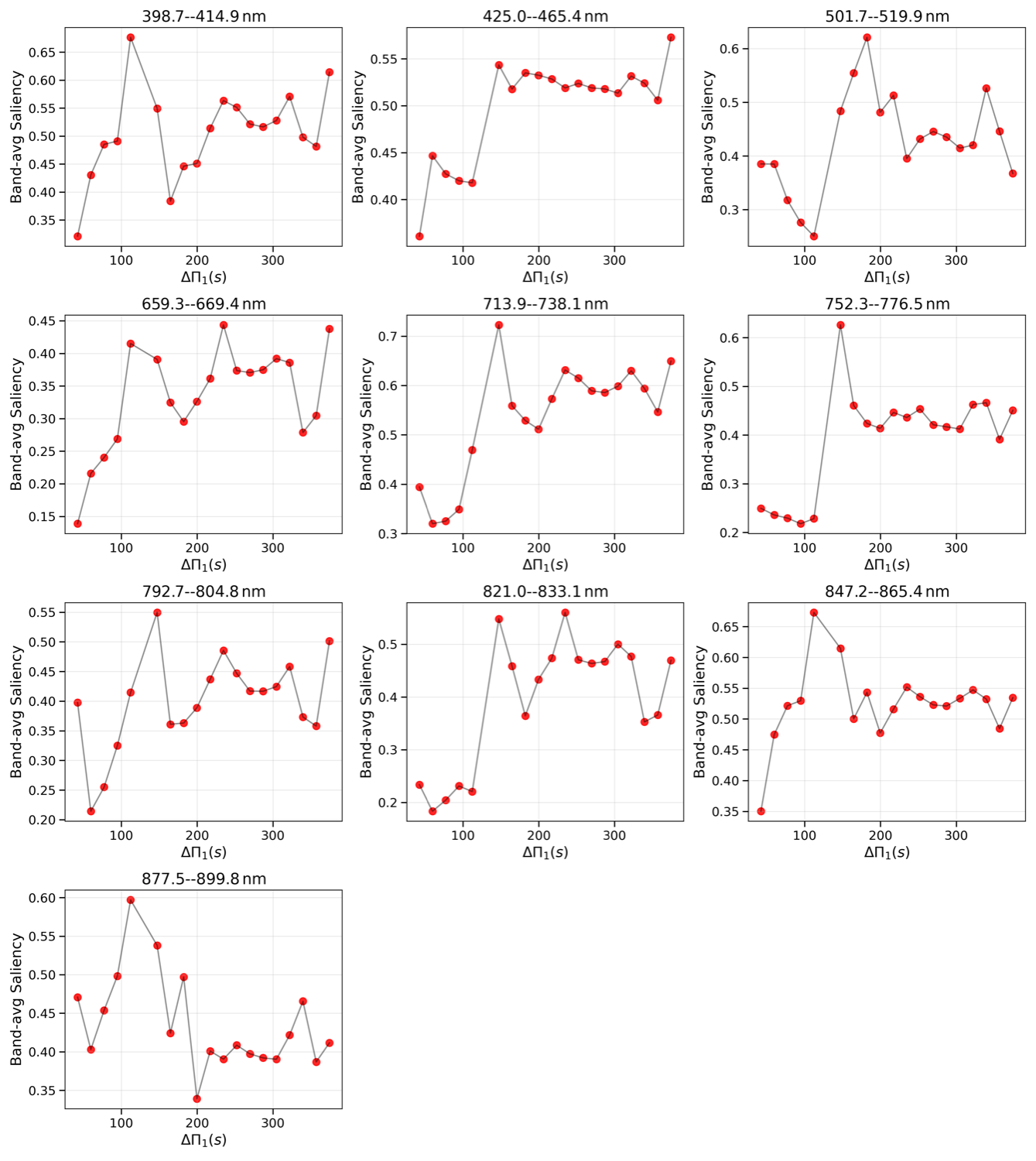}
    \caption{Average normalized saliency scores as a function of $\Delta\Pi_1$ for selected wavelength bands in the $\Delta\Pi_1$ model. The $\Delta\Pi_1$ range is divided into 20 bins, and saliency scores are averaged within each bin for bands with normalized saliency values above the adopted threshold.}
    \label{fig:saliency_trends_dpi}
\end{figure*}

For both the $\Delta\nu$ and $\Delta\Pi_1$ models, we divided the respective parameter ranges into 20 bins and computed the average normalized saliency scores across all wavelength bands, retaining only those bands with normalized saliency values above a specified threshold. Our aim is to assess whether different wavelength bands exhibit varying sensitivities across the $\Delta\nu$ and $\Delta\Pi_1$ ranges. As shown in Figures~\ref{fig:saliency_dnu} and \ref{fig:saliency_trends_dpi}, several bands display systematic trends, including a steady increase in saliency with increasing $\Delta\nu$ in the $372.4$--$374.5~\mathrm{nm}$ band.

\subsection{Comparison with External Catalogs}

To assess the robustness of our asteroseismic inferences from Gaia XP spectra, we compare our $\Delta \nu$ and $\Delta \Pi_1$ predictions with independent measurements from other catalogs. Stars in our training and test sets are excluded to ensure a fair comparison.
We test our inferred results against three catalogs: (a) The Complete Kepler Evolutionary Catalog of Red Giants by \citet{2025A&A...697A.165V}, (b) Asteroseismic measurements on Kepler Giants by \citet{dhanpal2022measuring} using deep-learning, and (c) Determination of $\Delta\nu$ and $\Delta\Pi_{1}$ from LAMOST spectra by \cite{wang2023precise}.

\subsubsection{The Complete Kepler Catalog}

For the purpose of cross-matching, we adopt the \textit{Kepler} red-giant evolutionary status catalog of \citet{2025A&A...697A.165V}.  
The catalog analyzed 30,337 giants and provided consensus classifications for 18,784 stars (11,387 RGB/AGB and 7,397 RC), including 11,516 with APOGEE spectra.

The evolutionary status was determined using six independent asteroseismic techniques - based on mixed-mode period spacings, pressure-mode patterns, autocorrelation, and machine learning - and agrees with spectroscopy at the $\sim$95\% level.  
A sharp lower boundary for RCs was identified at $\log g = 2.99 \pm 0.01$ \texttt{dex}, while the distinction between luminous RGB and AGB stars remains more uncertain. This catalog represents the most complete and precise evolutionary status map for \textit{Kepler} red giants to date.

\begin{figure}
    \centering
    \includegraphics[width=0.50\textwidth, height=0.36\textheight]{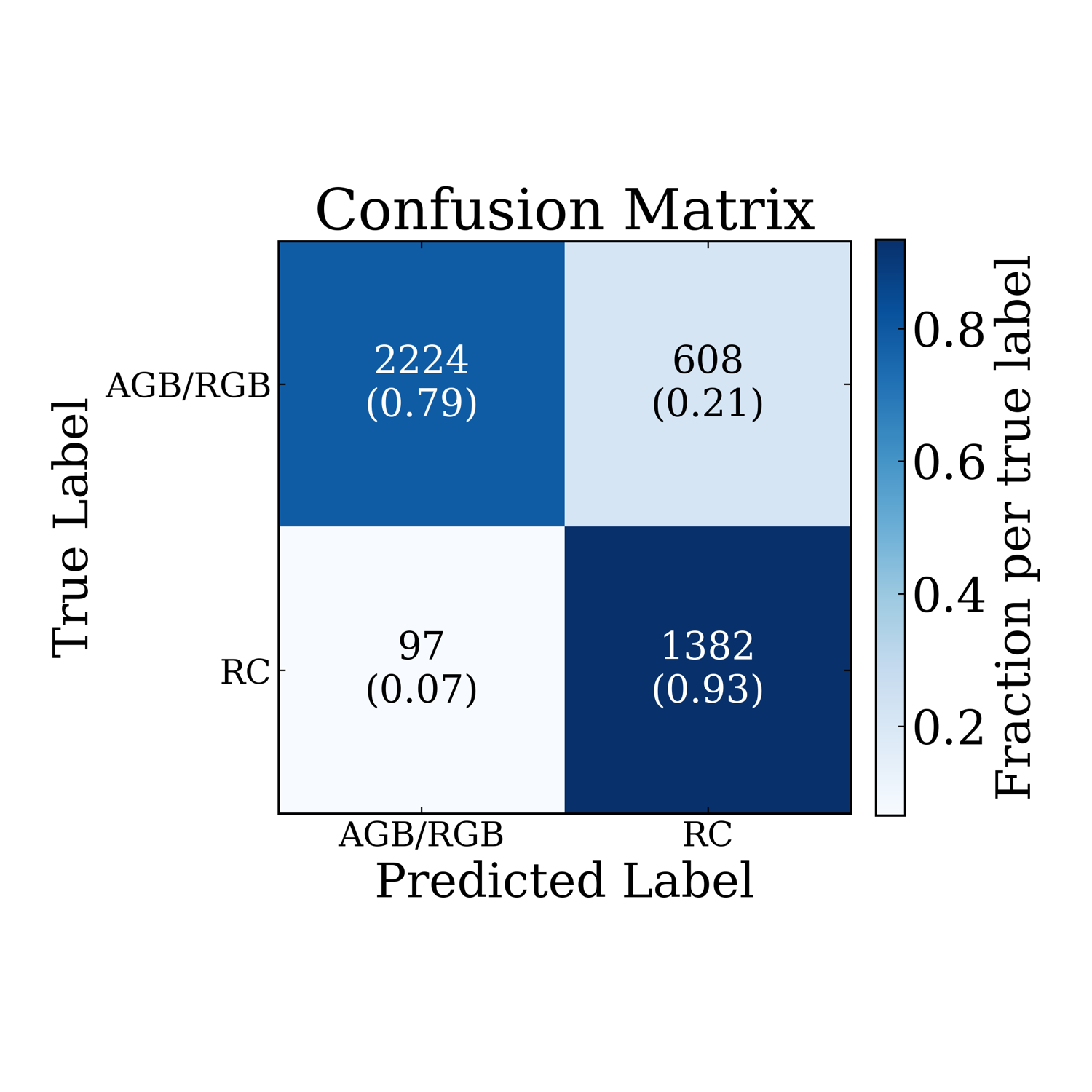}
    \caption{Confusion matrix comparing our $\Delta\Pi_{\rm Gaia}$-based classification with the evolutionary states reported in the \citet{2025A&A...697A.165V} catalog for 4311 cross-matched Gaia DR3 red giants. Our method recovers 93\% of RC stars and $\sim$80\% of RGB stars, yielding an overall agreement of 84\%, demonstrating a good discriminative ability.}
    \label{fig:vrard_rgb_rc}
\end{figure}

 Cross-matching our results with the \citet{2025A&A...697A.165V} catalog, we identified 4311 stars in our final inferred sample (outside the training range) for which an evolutionary stage (RGB/AGB or RC) is provided with certainty. We compared our classification based on the simple criterion $\Delta \Pi_{\rm Gaia} \geq 150$\,s, with their determinations. Overall, 84\% of the inferences show agreement between the two classifications. As shown in Figure~\ref{fig:vrard_rgb_rc}, our method correctly identifies 93\% of their red clump stars, while the corresponding fraction for red giant branch stars is $\sim$80\%.

\begin{table}
\centering
\renewcommand{\arraystretch}{1.2}
\begin{tabular}{|c c c|}
\hline
\textbf{Parameter pair} & \textbf{\textit{\rm{accuracy}90}} & \textbf{\textit{\rm{accuracy}80}} \\ \hline
$\Delta \nu_{\rm{Gaia}}$ vs $\Delta \nu_{p}$  & 35\% & 63\% \\ 
$\Delta \nu_{\rm{LAMOST}}$ vs $\Delta \nu_{p}$ & 50\% & 70\% \\ 
$\Delta \nu_{\rm{Gaia}}$ vs $\Delta \nu_{\rm{LAMOST}}$ & 71\% & 86\% \\ \hline
$\Delta \Pi_{\rm{Gaia}}$ vs $\Delta \Pi_{p}$  & 38\% & 65\% \\ 
$\Delta \Pi_{\rm{LAMOST}}$ vs $\Delta \Pi_{p}$ & 60\% & 76\% \\ 
$\Delta \Pi_{\rm{Gaia}}$ vs $\Delta \Pi_{\rm{LAMOST}}$ & 44\% & 67\% \\ \hline
\end{tabular}
\caption{Comparison between different catalogs (in terms of best values). The parameters used here are defined in Sections \ref{sec:SampleSelection}. $\Delta\nu_{\rm{Gaia}}$ and $\Delta\Pi_{\rm{Gaia}}$ are predictions of $\Delta\nu$ and $\Delta\Pi_{1}$ inferred by the ML model developed in this work.}
\label{table:matching}
\end{table}

\subsubsection{Comparison with Dhanpal et al. (Kepler)}

For cross-matching $\Delta\nu$ and $\Delta\Pi_{1}$ values, we adopt the catalog of \citet{dhanpal2022measuring}, which provides seismic parameters for \textit{Kepler} red giants. Large frequency separations (\(\Delta\nu\)) and period spacings (\(\Delta\Pi_1\)) were derived using a convolutional neural network trained on synthetic oscillation spectra based on asymptotic theory, incorporating p- and mixed-mode patterns with rotational effects. The method was validated against several thousand \textit{Kepler} stars and found to be in close agreement with published measurements.

A direct comparison shows that our $\Delta \nu_{\rm Gaia}$ are broadly consistent with $\Delta \nu_p$ from \citet{dhanpal2022measuring} (Figure~\ref{fig:dnu_match}(a)). We note, however, a systematic offset (as a vertical feature) in the high-$\Delta \nu$ regime ($\gtrsim 18.5$ $\mu$Hz). This can be understood as a consequence of the limited number of training samples in this range: only 24 stars ($\sim$0.1\% of the training set) lie above this threshold, although the training data extend to $\sim$19.3 $\mu$Hz. In this regime, the model is therefore only weakly constrained, and modest deviations are expected.

Likewise, $\Delta\Pi_{Gaia}$ are broadly consistent with $\Delta\Pi_p$ (Figure~\ref{fig:dpi_match} (a)), with some deviations. In particular, the linear feature along $\Delta\Pi_{\rm Gaia}$ observed over 70 s < $\Delta\Pi_p$ < 90 s appears to be related to the characteristics of the training dataset used for training. The training data are inherently imbalanced, with $\sim$2,000 stars having $\Delta\Pi_1$ < 150 s and $\sim$4,000 stars above this threshold. Within the $\Delta\Pi_1$ < 100 s regime, $\sim$180 out of 1,462 training stars ($\sim$12.3\%) fall in this interval, while a comparable fraction is recovered in Figure~\ref{fig:dpi_match}(a) (88 out of 665 stars, $\sim$13.2\%). The close agreement suggests that this feature likely reflects the underlying training distribution, although additional factors may also contribute.

\subsubsection{Comparison with LAMOST}

\begin{figure*}
    \centering
    \begin{subfigure}[t]{0.42\textwidth}
        \centering
        \includegraphics[width=\textwidth]{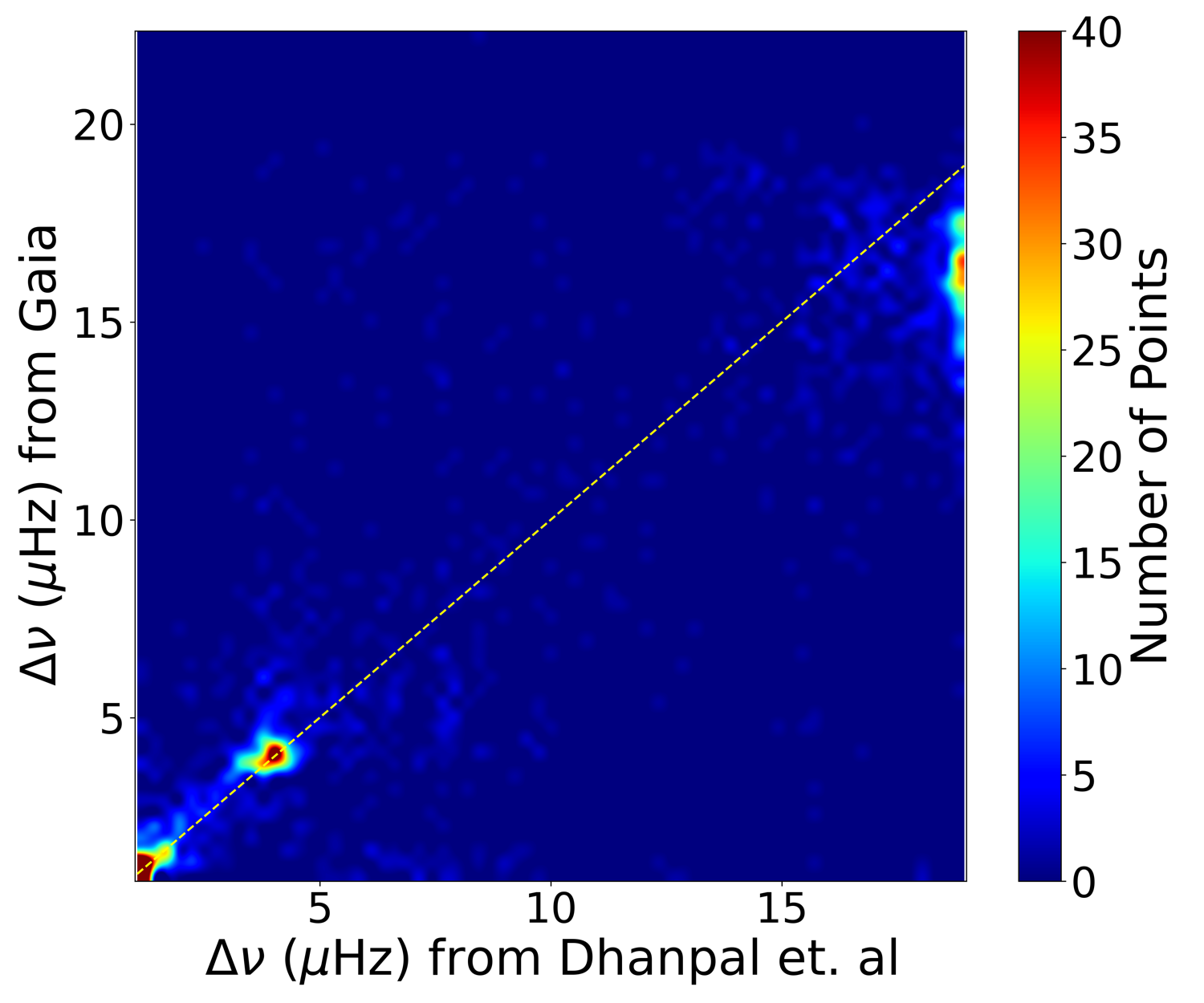}
        \caption{Comparison of $\Delta \nu_p$ values from \citet{dhanpal2022measuring} with $\Delta \nu_{\rm{Gaia}}$ predicted from Gaia XP spectra. The color map shows point density, and Gaussian contours highlight the distribution.}
    \end{subfigure}
    \hfill
    \begin{subfigure}[t]{0.42\textwidth}
        \centering
        \includegraphics[width=\textwidth]{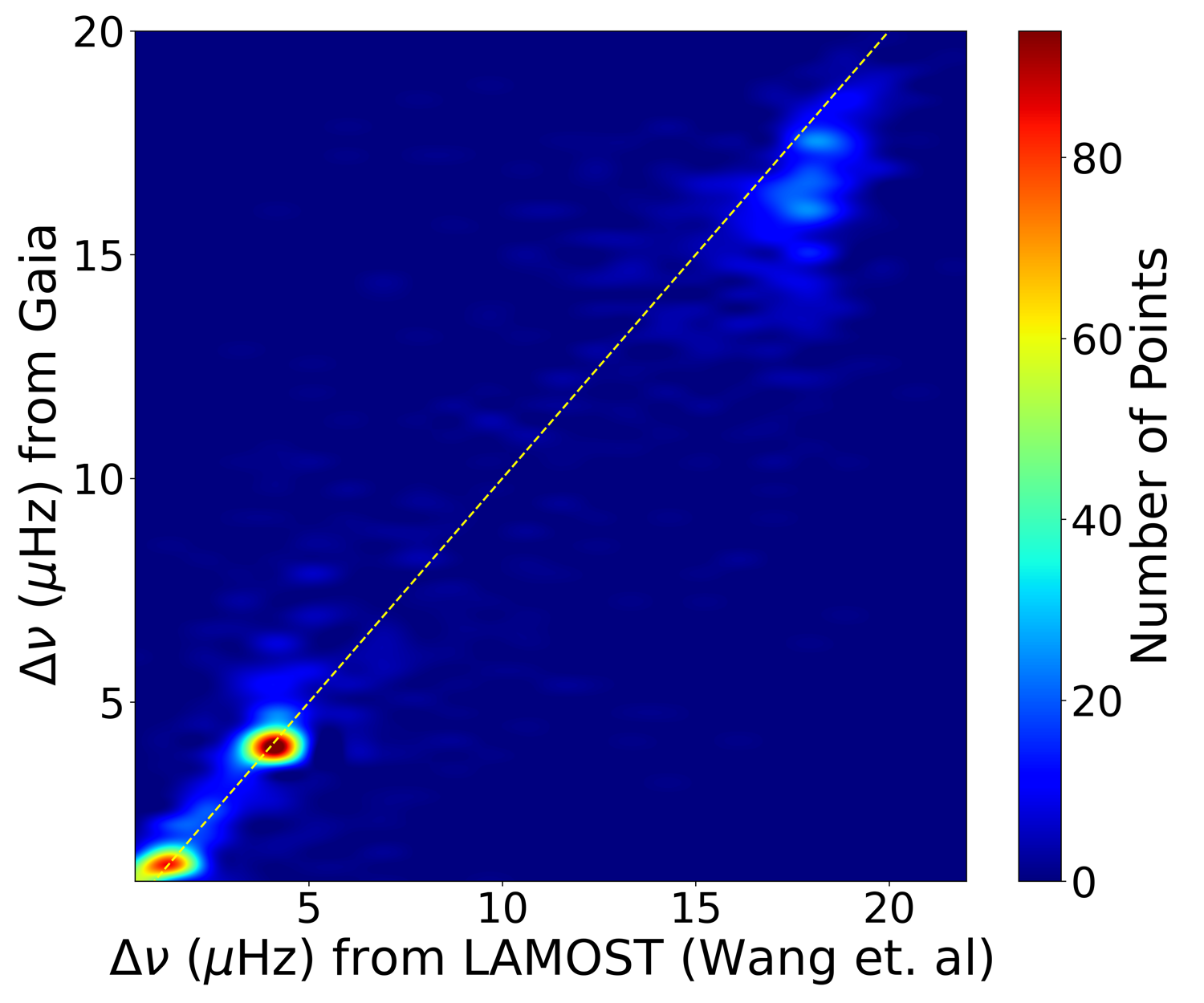}
        \caption{Top: Comparison of $\Delta \nu_{\rm{LAMOST}}$ predictions with \citet{wang2023precise}. Bottom: Comparison of $\Delta \nu_{\rm{LAMOST}}$ from \citet{wang2023precise} with $\Delta \nu_{\rm{Gaia}}$ from XP spectra. Gaia predictions align well with power-spectral inferences, though some scatter exists.}
    \end{subfigure}
    \caption{Comparisons of asteroseismic $\Delta \nu$ values derived from Gaia XP spectra with those from independent power spectrum analyses and LAMOST spectra.}
    \label{fig:dnu_match}
\end{figure*}

\begin{figure*}
    \centering
    \begin{subfigure}[t]{0.44\textwidth}
        \centering
        \includegraphics[width=\textwidth]{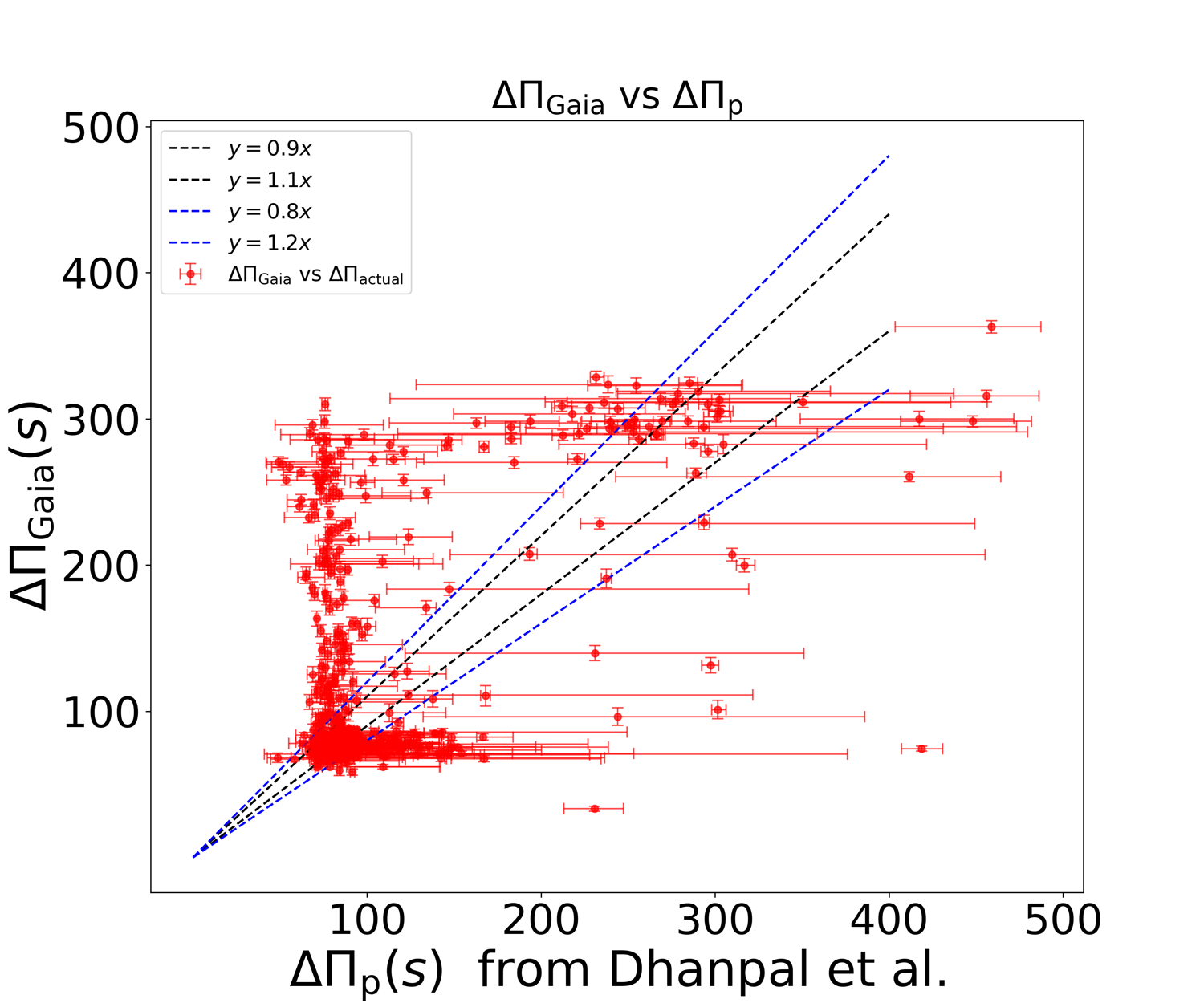}
        \caption{Comparison of power spectral inferences of $\Delta \Pi_{1}$ with Gaia XP predictions.}
    \end{subfigure}
    \hfill
    \begin{subfigure}[t]{0.44\textwidth}
        \centering
        \includegraphics[width=\textwidth]{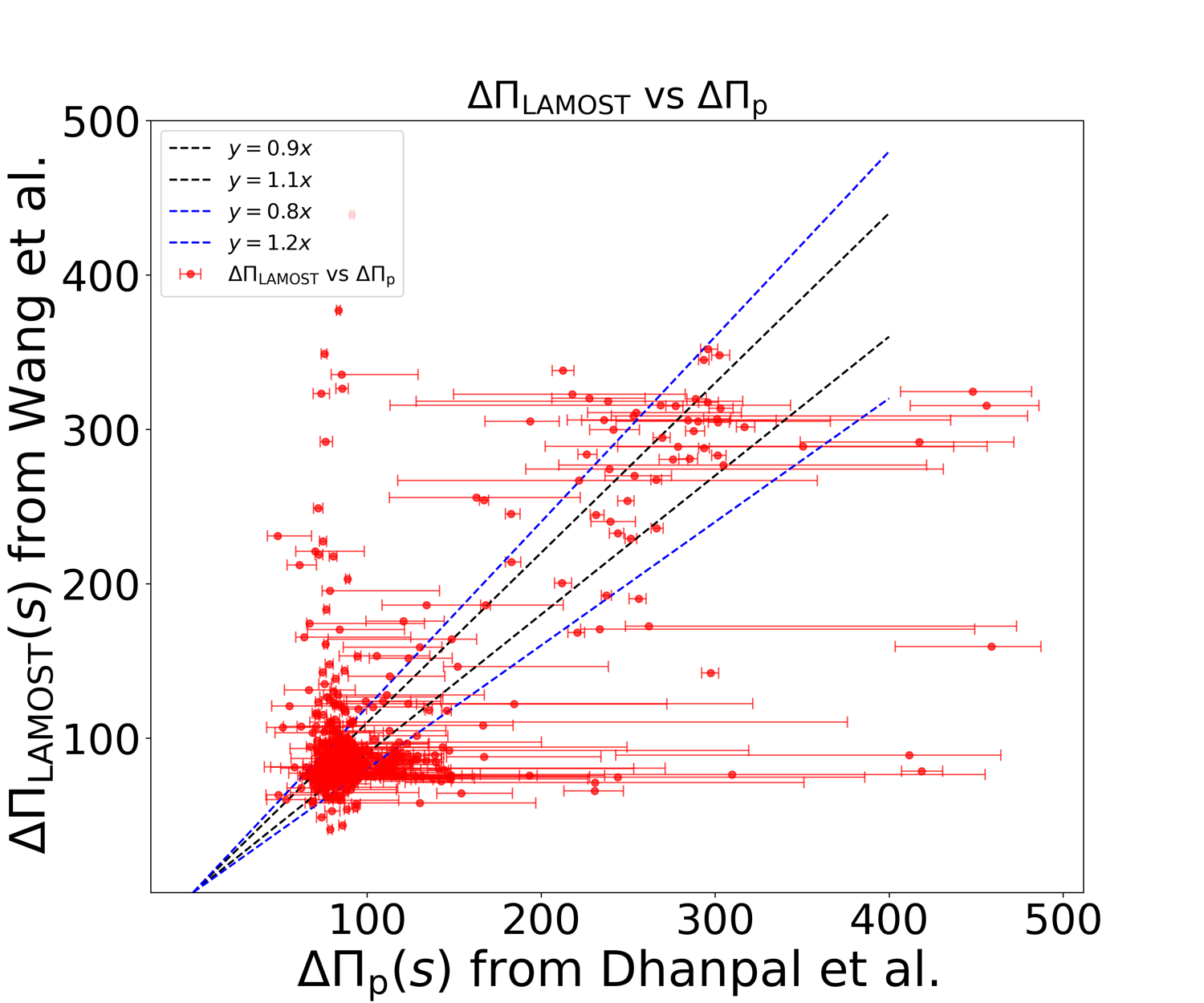}
        \caption{Comparison of power spectral inferences of $\Delta \Pi_{1}$ with LAMOST predictions.}
    \end{subfigure}
    \caption{Comparison of power spectral inferences of $\Delta \Pi_{1}$ with Gaia and LAMOST-based predictions.}
    \label{fig:dpi_match}
\end{figure*}

For cross-matching of $\Delta\nu$ and $\Delta\Pi_{1}$, we use the catalog of \citet{wang2023precise}, who employed the same asteroseismic training sample as in our work but applied it to LAMOST spectra at $R \sim 1800$. In their approach, both $\Delta\nu$ and $\Delta\Pi_{1}$ were inferred directly from the spectra using a data-driven model, yielding homogeneous seismic parameters for a large sample of giants. We further validate our Gaia-based inferences against these LAMOST measurements: red giants common to Gaia and LAMOST show good agreement in both $\Delta\nu$ (Figure~\ref{fig:dnu_match}) and $\Delta\Pi_1$ (Figure~\ref{fig:dpi_match}), consistent with the trends observed in the \textit{Kepler} comparison. Quantitative metrics are summarized in Table~\ref{table:matching}, confirming the robustness of our Gaia-derived asteroseismic parameters.

We note, however, that at higher $\Delta \nu$ values, a difference of approximately 5–10\% between $\Delta \nu_{\rm LAMOST}$ and $\Delta \nu_{\rm Gaia}$ is not unexpected. This likely reflects a combination of factors, including differences in the underlying training distributions, the size of the training samples - $\sim$1,800 stars in \citet{wang2023precise} compared to $\sim$15,000 for $\Delta \nu$ and $\sim$4,800 for $\Delta\Pi_1$ in this work - as well as differences in training strategies and spectral data quality.

\subsection{Results for new Gaia DR3 red giants}
\label{sec: results_4.3}
\begin{figure*}
    \centering
    \begin{subfigure}[t]{0.7\textwidth}
        \centering
        \includegraphics[width=\textwidth]{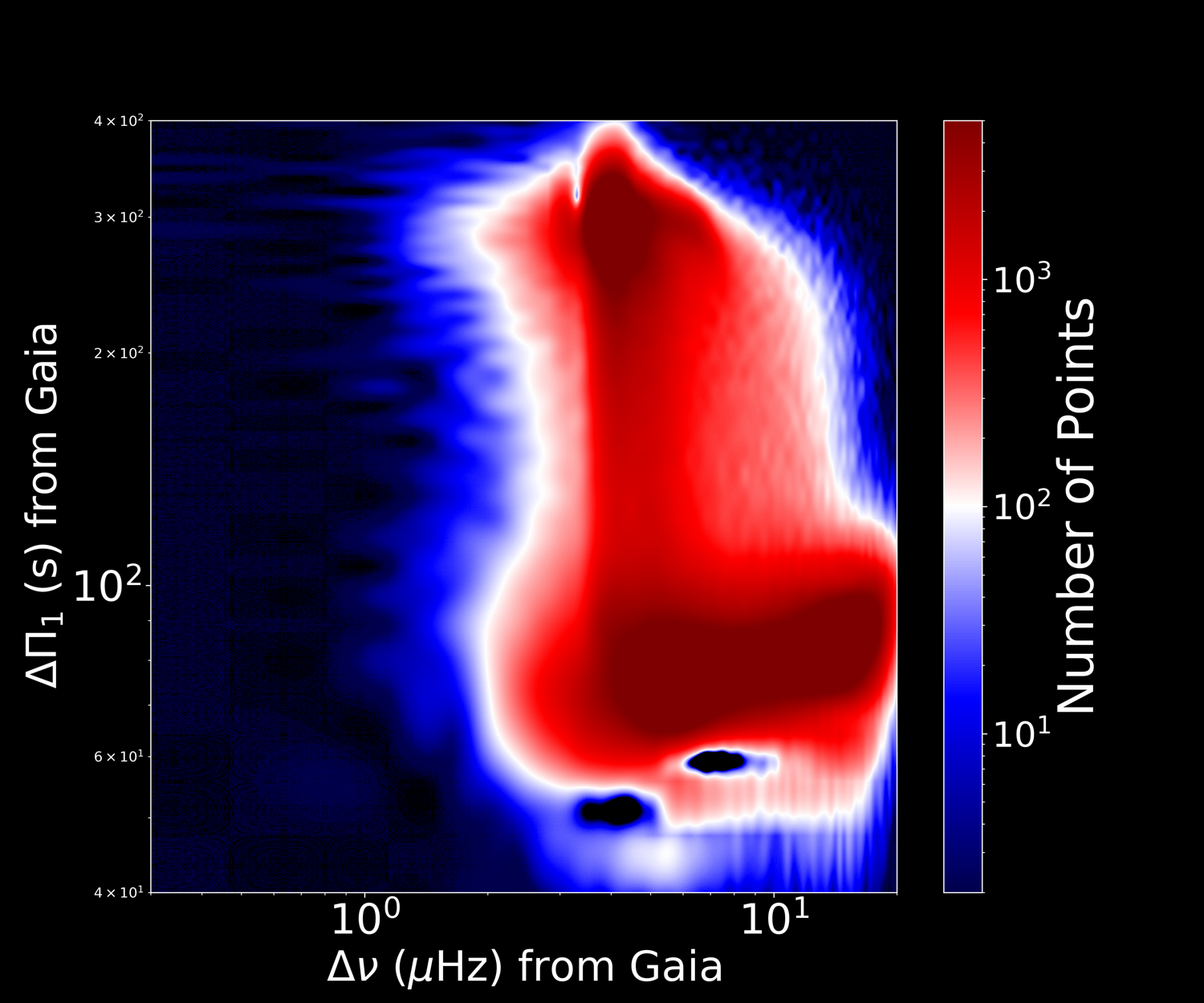}
        \caption{$\Delta \nu$–$\Delta \Pi_{1}$ plot for 2,557,881 red giants (selected using the cuts mentioned in \ref{sec : Predcitions_params}), derived from Gaia XP. The horizontal feature extends up to $\Delta\nu \sim 1 \, \mu$Hz for the 13,812 stars classified as RCs \citep[common between our classification and that of][]{Bovy_2014}, although it is not clearly visible here.}
    \end{subfigure}
    
    \vspace{1em}
    
    \begin{subfigure}[t]{0.7\textwidth}
        \centering
        \includegraphics[width=\textwidth]{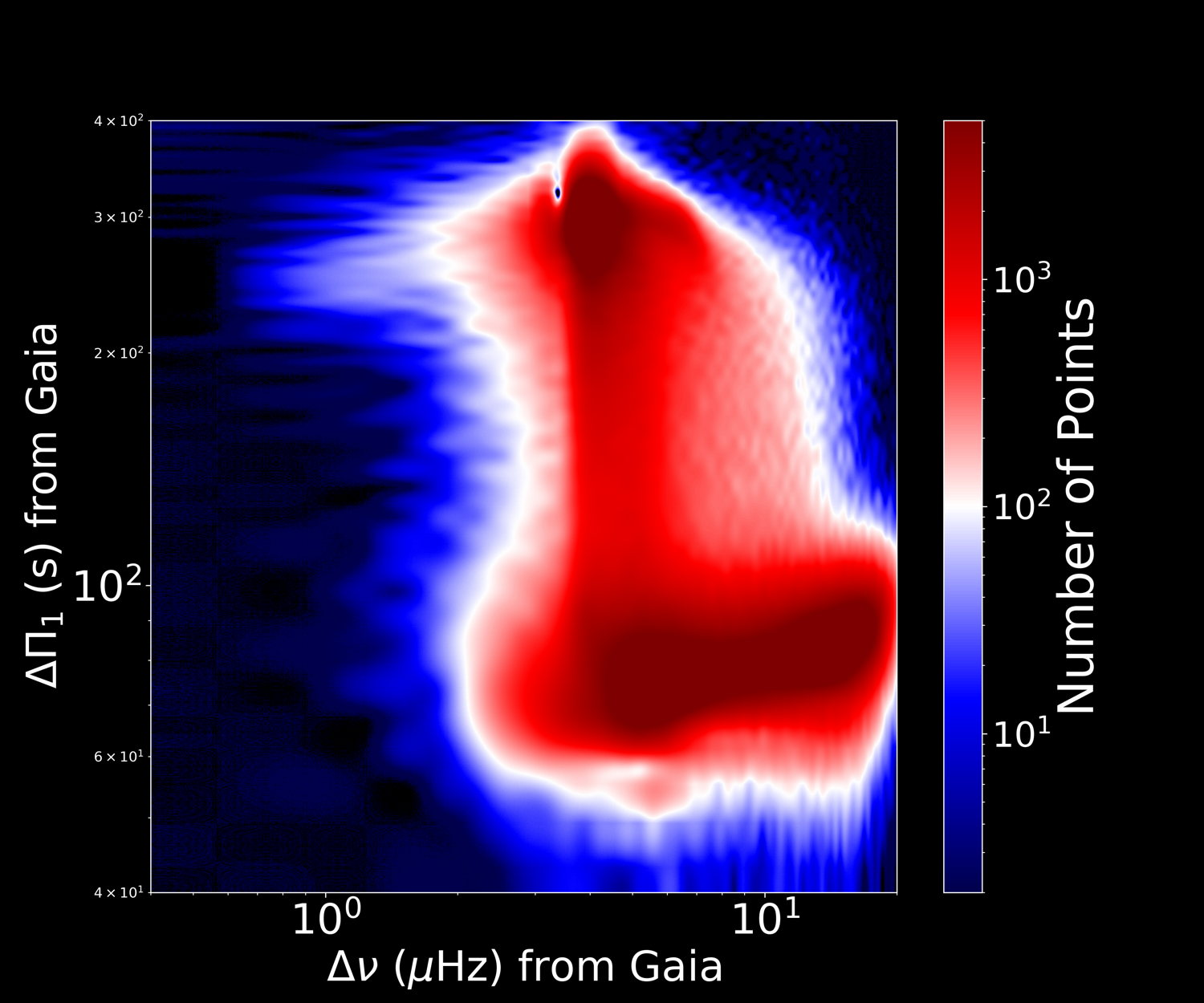}
        \caption{Same plot as (a), but now including lower metallicity stars . The arm-like structure becomes more prominent in this version.}
    \end{subfigure}
    
    \caption{Global asteroseismic parameter distributions derived from Gaia XP spectra.}
    \label{fig:dnu_dpi_gaia}
\end{figure*}

Applying the $T_{\rm{eff}}$, $\rm{[M/H]}$, $\log g$ and reddening criteria (refer Section \ref{sec : Predcitions_params}) to the full sample of 17,558,141 red giants, we are left with 3,073,850 stars for $\Delta\nu$ and 2,557,881 stars for $\Delta\Pi_{1}$ \citep{Andraeetal2023a}. These cuts are essential to remove biased predictions in the final Gaia DR3 sample. We further ensured these targets have good quality (\texttt{phot\_bp\_mean\_flux\_over\_error} $>$ 10 and \texttt{phot\_rp\_mean\_flux\_over\_error} $>$ 10) Gaia XP sampled spectra (\texttt{has\_xp\_sampled = 'true'}). 
Finally we obtain 2,557,881 red giant Gaia XP spectra common to both $\Delta\nu$ and $\Delta\Pi_{1}$, and using our neural-network models, we predicted the values of $\Delta \nu$ and $\Delta \Pi_{1}$ for these stars. The results are displayed in Figure \ref{fig:dnu_dpi_gaia}. Notably, despite the overall reliability of our predictions, we observed 1290 instances of $\Delta \Pi_{1}$ predictions which are outside the training range. 
However, these represent only approximately 0.05\% of the total data points, underscoring the high retrieval accuracy of our model. The complete dataset can be accessed as the supplementary material attached to this paper.

We detect an arm-like feature in the RC region of our final $\Delta\nu$–$\Delta\Pi_{1}$ diagram, specifically for $\Delta\nu \leq 3.5$ $\mu$Hz (Figure \ref{fig:dnu_dpi_gaia}b). This feature is centered around $\Delta\Pi_{1} = 290$ s for stars with metallicities in the range of training data (Figure \ref{fig:dnu_dpi_gaia}a), and has a positive slope when metallicities lower than those in the training data (Figure \ref{fig:dnu_dpi_gaia}b) are also included. Approximately 1.3\% of the 2,557,881 stars exhibit this unexpected trend around $\Delta\Pi_{1} = 290$ s. Adjacent to these arm-like features is a very prominent kink and dip (valley)-like feature between $\Delta\nu = 3.0$ and $3.5 \: \mu\rm{Hz}$.
Despite the presence of very few of stars ($\sim$ 0.6\%) in this region of the $\Delta\nu-\Delta\Pi_{1}$ diagram, no distinct feature has been reported in previous studies like those of \cite{2014A&A...572L...5M} and \cite{Vrard_2016}.
In particular, our training sample provides sufficient coverage in this parameter space, incorporating $\Delta\nu$ values from \citet{Yu_2018} and $\Delta\Pi_{1}$ values from \citet{Vrard_2016}. Since our models independently predict these parameters, the presence of stars in this region is not entirely unexpected. Additionally, obtaining independent confirmation of these measurements, e.g., through asteroseismology, requires fitting to the observed oscillation spectra at these low values of $\Delta\nu$, a challenging proposition due to the high density of modes. It is worth noting that there is also a vacuole-like feature present just below the RGB stars. However, only 0.41\% of the total training samples of $\Delta\Pi_1$ lie in this range, which makes these inferences less reliable.
Moreover when low metallicity stars are taken into consideration, the vacuole-like feature is absent whereas the arm-and-valley like features become more pronounced (see next subsection \ref{subsec: results_4.4}).

A close candidate for the explanation of the arm-like feature could be the recent theoretical study by \cite{Capelo2023}, where they use MESA models of a $1.6 M_{\odot}$ star \citep[with metallicity 0.01834, whose relative abundances are given by][]{GS98} to show that after the Helium flash, the star follows a track with a small positive slope on the log-log plot of the $\Delta\nu$-$\Delta\Pi_{1}$ diagram (having $\Delta\Pi_{1} \in [300 \: \rm{s}, 400 \: \rm{s}]$ for $\Delta\nu \in [1 \: \mu\rm{Hz}, 3 \: \mu\rm{Hz}]$), succeeded by a kink near $\Delta\nu \approx 3\;\mu\rm{Hz}$ before reaching the primary red-clump zone.

In their model, neutrino emission from plasmon decay is suppressed during the Helium flash and subsequent sub-flashes are not required to lift the Helium core out of degeneracy. The feature reported by them in the aforementioned case is not seen in the plot for the case where the channel for thermal transport through plasmon decay neutrinos is permitted. Whether this mechanism fully explains the feature in Gaia data remains an open question.  

\begin{figure}
    \centering
    \includegraphics[width=0.50\textwidth, height=0.30\textheight]{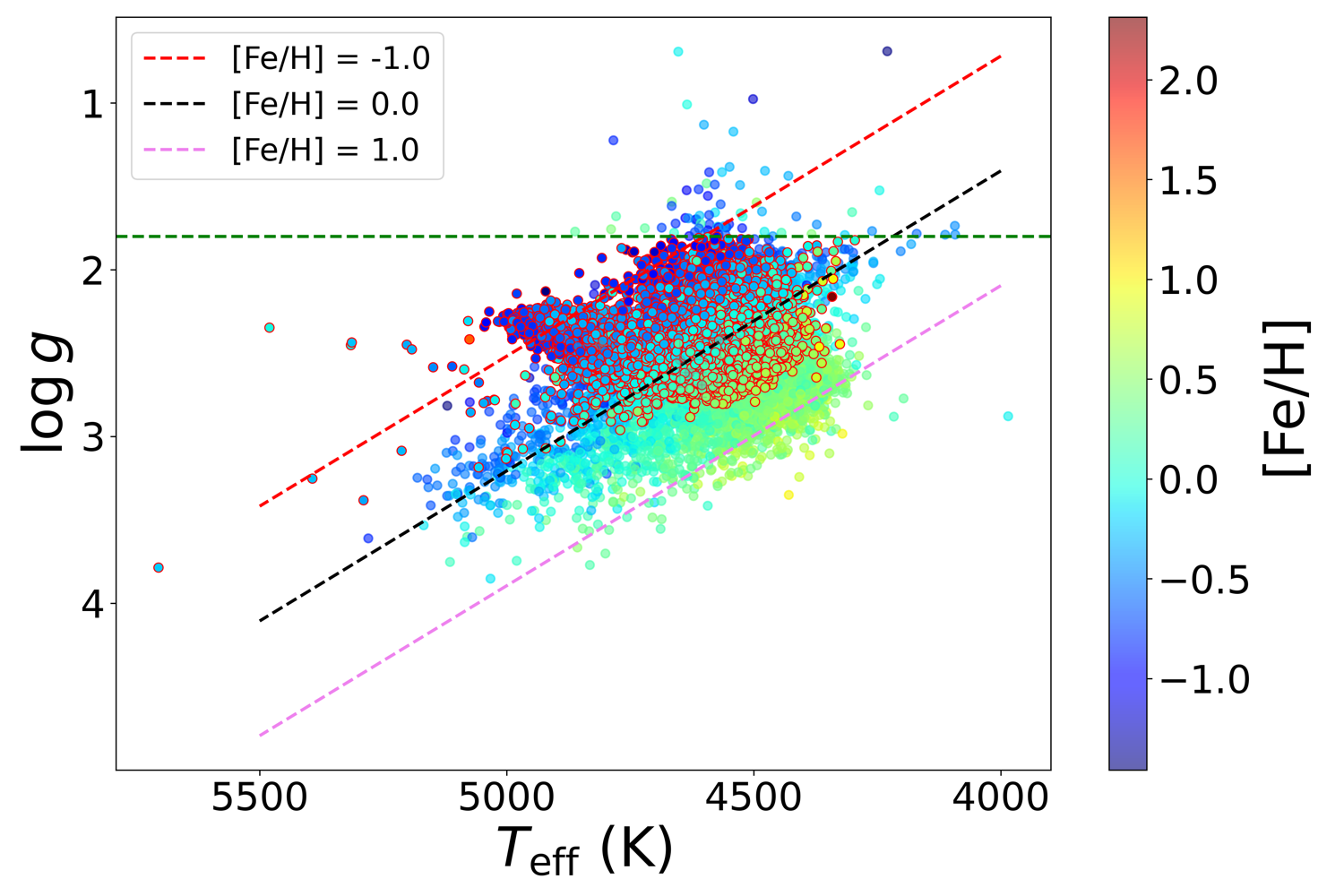}
    \caption{The red circled stars are classified as RCs based on the selection criteria of \citet{Bovy_2014}. The red, black, and violet lines represent [Fe/H]-dependent cuts that separate RCs from RGBs. The green horizontal dashed line at $\log g = 1.8$ marks the lower bound for RC classification. Out of 35,515 stars, 13,812 satisfy the RC criteria.}
    \label{fig:low_logg_teff}
\end{figure}

To assess whether these stars align also with the classical RC classification, we first selected a sample of 35,515 stars with $\Delta\Pi_1$ within 290 $\pm$ 40 s and $\Delta\nu < 3.5$ $\mu$Hz from our model predictions. We then applied the criteria of \citet{Bovy_2014} (although in their Figure 1, a significant number of stars remain unclassified and some contamination persists even after applying the selection cuts) using $\log g$, $T_{\rm eff}$, and $[\rm{Fe/H}]$ values from \citet{Zhang_2023}, which were derived through forward modeling of Gaia XP spectra. 
Out of these, 13,812 stars satisfy their RC classification criteria. Further cuts (6, 7 and 8 in \citet{Bovy_2014}) using metallicities from \citet{Andraeetal2023a} and $(J - K_s)_0$ colors from 2MASS photometry suggests that 5,619 stars belong to the primary red clump (PRC). Despite the
 presence of these stars in the low-$\Delta\nu$ regime, the estimated purity of this final sample remains high ($\sim$ 93\%) \citep{Bovy_2014, huang2015metallicitygradientsgalacticdisk}, suggesting that the model is potentially capturing realistic patterns. However,  further investigation is needed in order to verify these results. The $\log g$–$T_{\rm eff}$ distribution for this sample is shown in Figure \ref{fig:low_logg_teff}, where the red-circled stars follow the expected RC characteristics provided by \citet{Bovy_2014}.

Although asteroseismically measuring $\Delta\nu\lesssim 4~\mu$Hz, i.e., at the lower end, is challenging due to the high density of oscillation modes, the presence of oscillation power itself can still be assessed. Based on our selection criteria and the empirical scaling relation between $\Delta\nu$ and $\nu_{\mathrm{max}}$ , which is valid in this regime \citep{Stello_2009}, the expected oscillation power should peak around $\nu_{\mathrm{max}} \sim 35.6\,\mu\mathrm{Hz}$ or below.

To verify this expectation, we analyzed TESS-SPOC and \textit{Kepler} light curves for the available targets and computed their Lomb-Scargle periodograms (LSPs), using the \texttt{Lightkurve} package \citep{2018ascl.soft12013L}, and examined the frequency range of the oscillation power. The detection of acoustic-mode envelopes in their LSPs with peaks in the anticipated $\nu_{\rm{max}}$ regime also provides an independent consistency check on the reliability of the $\Delta\nu$ values inferred from Gaia XP spectra using our machine-learning framework.

\begin{figure*}
    \centering
    \includegraphics[width=\textwidth]{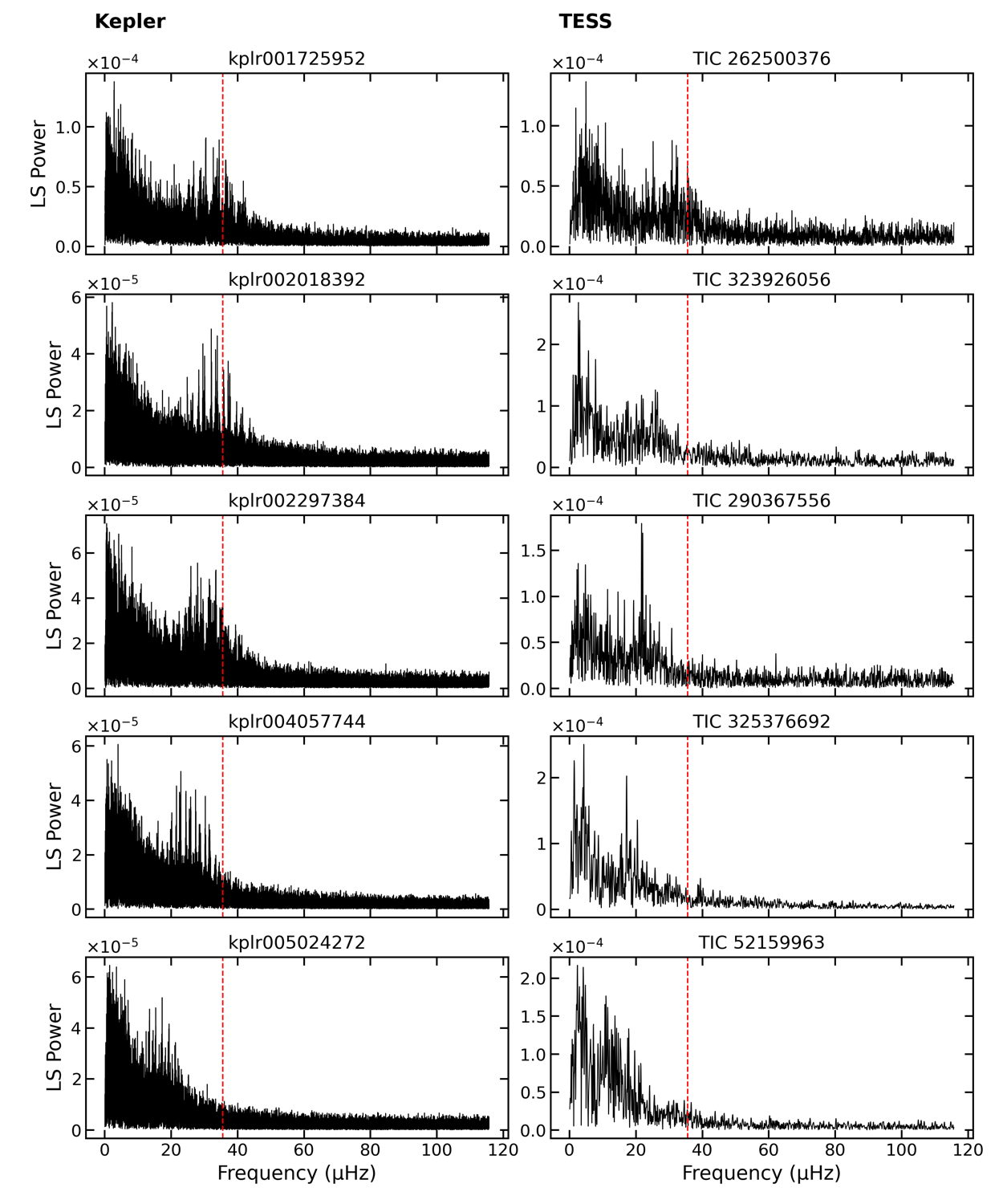}
    \caption{Example of power spectra of few red giants which have $\Delta\nu \le 3.5 $ $\mu\rm{Hz}$ and $\Delta\Pi_1$ in between 290 s $\pm$ 40 s from our Gaia XP inference. The dotted vertical red line indicates the upper limit of the permitted value of $\nu_{\mathrm{max}}$ corresponding to $\Delta\nu \le 3.5~\mu\mathrm{Hz}$.
    }
    \label{fig:power_spectra}
\end{figure*}

Figure~\ref{fig:power_spectra} presents the power spectra of ten representative red giant stars selected from the $\Delta\nu \leq 3.5\,\mu\mathrm{Hz}$ clump region (additional examples are shown in Appendix B). In all cases, the excess power is observed at low frequencies, consistent with oscillations in the low-$\nu_{\mathrm{max}}$ regime, thereby supporting the interpretation that these stars occupy the expected evolutionary phase.

\subsection{Machine inferences on metal-poor Gaia red giants}
\label{subsec: results_4.4}

Our training sample of \textit{Kepler}-\textit{Gaia} red giants predominantly comprises stars with metallicities ranging from $-0.7$ \texttt{dex} to $0.4$ \texttt{dex}, with only a sparse representation of stars below $-1.0$ \texttt{dex}. To assess the performance of our machine learning model in this lower-metallicity regime, we selected red giants from \citet{Andraeetal2023a} with $[\rm{M/H}] < -1.0$ \texttt{dex}, while ensuring that their effective temperatures ($\rm{T}_{\rm{eff}}$) and surface gravities ($\log g$) remained within the training range. As shown in Figure \ref{fig:dnu_dpi_gaia}(b), incorporating these metal-poor stars results in a more pronounced arm-like feature at low $\Delta\nu$ values within the red-clump region. At this stage, our machine-learning framework does not allow us to definitively determine the origin of this structure — it may be a genuine astrophysical effect or an artifact of the machine-learning model. Assessing the reliability of these predictions requires further investigation.

\section{Conclusion}

In this study, we have developed deep-learning models that use Gaia XP spectra to infer global asteroseismic parameters, $\Delta \nu$, $\nu_{\mathrm{max}}$, and $\Delta \Pi_{1}$. Previous works have demonstrated that single-epoch spectra can be used to derive these parameters using data-driven approaches with APOGEE ($\mathcal{R} \sim 22{,}500$) and LAMOST ($\mathcal{R} \sim 1{,}800$) spectra. However, Gaia XP spectra have significantly lower resolution ($\mathcal{R}$ $\sim$ 15-85), with unresolved spectral features and blending of important diagnostic features. Several spectral regions known to be critical for distinguishing red clump (RC) and red giant branch (RGB) stars—such as Fe5270 (5248--5252~\AA), MgH \& Mg~I (5167--5191~\AA), Fe~I (4956--4960~\AA), Fe~I (4205--4209~\AA), Cr~I (5205--5213~\AA), and CN (4173--4178~\AA) \citep{2022MNRAS.512.1710H} —are either heavily blended or unresolved in Gaia XP spectra. Remarkably, despite these limitations, our results reveal that the Gaia XP spectra contain subtle but crucial details beyond elemental abundances that codify essential information relating to the internal structure of red giants. Using the saliency analysis, we also identify the spectral regions that are most informative for distinguishing RC stars from RGB stars. These wavelength intervals may contain underlying physical signatures, and future work can investigate the specific astrophysical processes imprinted in them.

Our models enable the inference of asteroseismic parameters for approximately 2.5 million red giants in Gaia DR3, representing the largest and most comprehensive sample of asteroseismic data for red giants to date. Furthermore, from our results, we also obtain a large sample of RCs which can be used as standard candles in order to map the Milky Way. 

We also identify a subset of red clump stars with $\Delta\nu \le 3.5~\mu\mathrm{Hz}$ which shows a distinct horizontal feature in the $\Delta\nu$-$\Delta\Pi_1$ plane. We further examined the evolutionary status of these red giants using the criteria proposed by \citet{Bovy_2014}, and found that a significant fraction of these stars are consistent with our classification. We note, however, that failure to satisfy these criteria does not necessarily preclude a star from being in the red clump.

For a subset of these targets, we analyzed the available \textit{TESS} and \textit{Kepler} light curves and computed their Lomb-Scargle periodograms, examples of which are presented in this work. These periodograms clearly show oscillation power extending well below the maximum possible value of $\nu_{\rm{max}}$ corresponding to $\Delta\nu = 3.5~\mu\mathrm{Hz}$ (considering a maximum deviation of $20\%$ around the scaled $\nu_{\rm{max}}$ from $\Delta\nu$), consistent with their low-$\Delta\nu$ nature.

Although certain works in literature may explain the physical processes involved in the existence of stars in the arm-like feature on the $\Delta\nu-\Delta\Pi_{1}$ diagram \citep{Capelo2023}, we emphasize that a more detailed investigation is required to robustly confirm the origin and persistence of this horizontal feature. Such an analysis is beyond the scope of the present work and will be pursued in a forthcoming study.

\begin{acknowledgements}
RB, SB, SMH and SD thank the anonymous reviewer for their constructive suggestions which have improved the quality of the work. We further acknowledge support from the Department of Atomic Energy, Government of India, under Project Identification No. RTI 4002. This research was supported in part by a generous donation (from the Murty Trust) aimed at enabling advances in astrophysics through the use of machine learning. Murty Trust, an initiative of the Murty Foundation, is a not-for-profit organisation dedicated to the preservation and celebration of culture, science, and knowledge systems born out of India. The Murty Trust is headed by Mrs. Sudha Murty and Mr. Rohan Murty.
\end{acknowledgements}

\bibliographystyle{aa}

\newpage

\begin{appendix}

\section{5-fold Cross-Validation}

\label{app:5foldXvalidation}
To ensure that our results are robust and not dependent on a particular subset of the data, we applied \emph{5-fold cross-validation}, a standard technique in machine learning \citep{BROWNE2000108}. The procedure is as follows:

\begin{enumerate}
    \item The dataset is divided into five equal subsets (folds).
    \item The model is trained on four folds and tested on the remaining fold.
    \item This process is repeated five times, each time using a different fold as the test set.
\end{enumerate}

This approach serves multiple purposes:

\begin{itemize}
    \item \textbf{Reduces overfitting}: By evaluating the model on multiple independent test sets, we ensure that it does not simply memorize the training data \citep{hastie2009elements}. 
    \item \textbf{Provides uncertainty estimates}: The standard deviation across folds indicates the variability in model performance, giving a measure of reliability \citep{Arlot_2010}.
    \item \textbf{Maximizes data utilization}: Every data point is used for both training and validation, making the best use of limited observational data.
\end{itemize}

The 5-fold cross-validation results of our chosen set of hyper-parameters is shown below in Fig. \ref{fig:cross_val_res}.

\begin{figure}[b]
    \centering
    \begin{subfigure}[t]{0.45\textwidth}
        \centering
        \includegraphics[width=\textwidth]{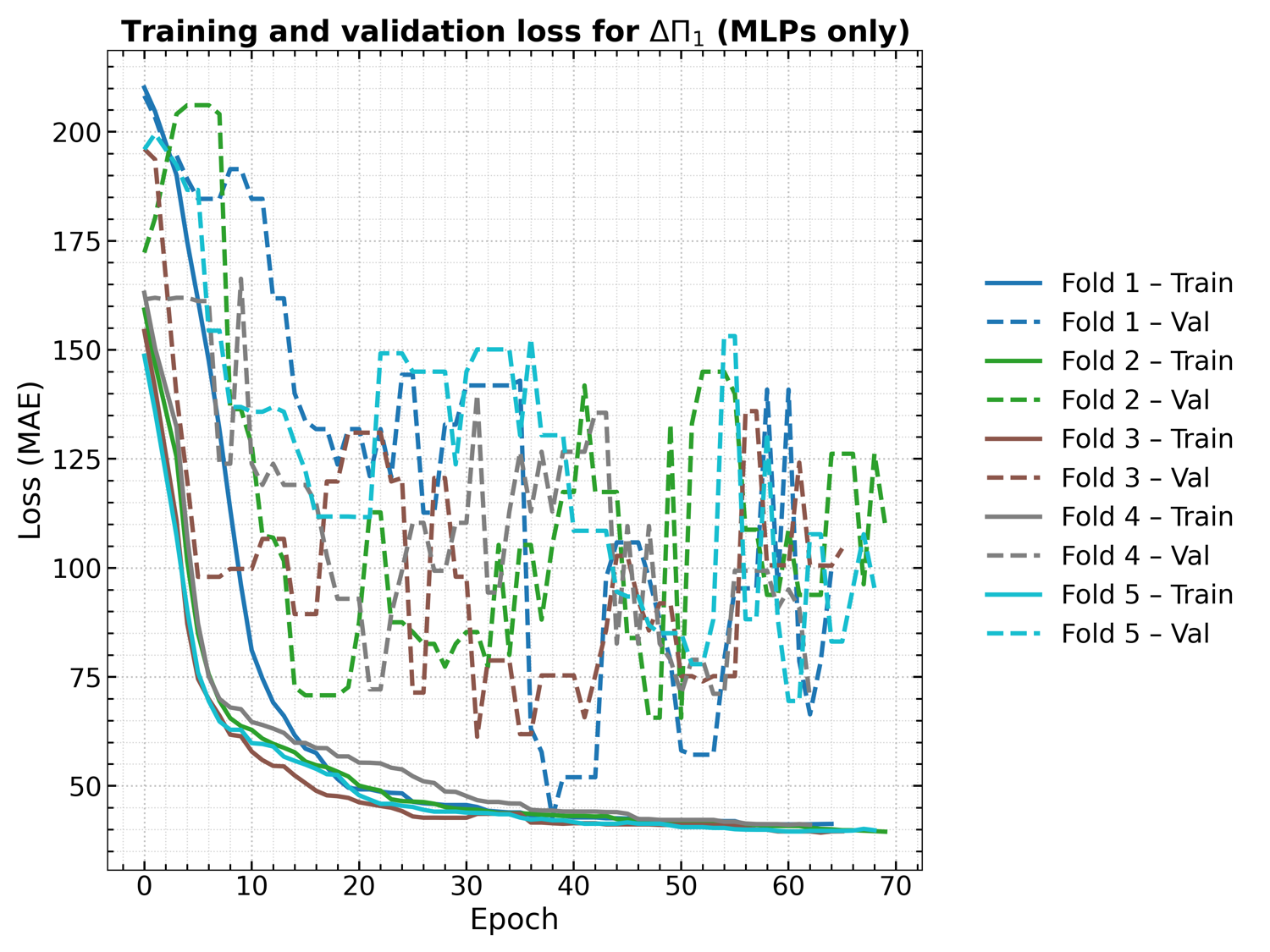}
        \caption{Cross-validation result of the $\Delta\Pi_1$ model that uses just MLPs. As can be seen, the generalization is poor.}
    \end{subfigure}
    \caption{Cross-validation results for a linear only-$\Delta\Pi_{1}$ model.}
    \label{fig:cross_val_res}
\end{figure}

\begin{figure*}
    \centering

    \begin{subfigure}[t]{0.4\textwidth}
        \centering
        \includegraphics[width=\textwidth]{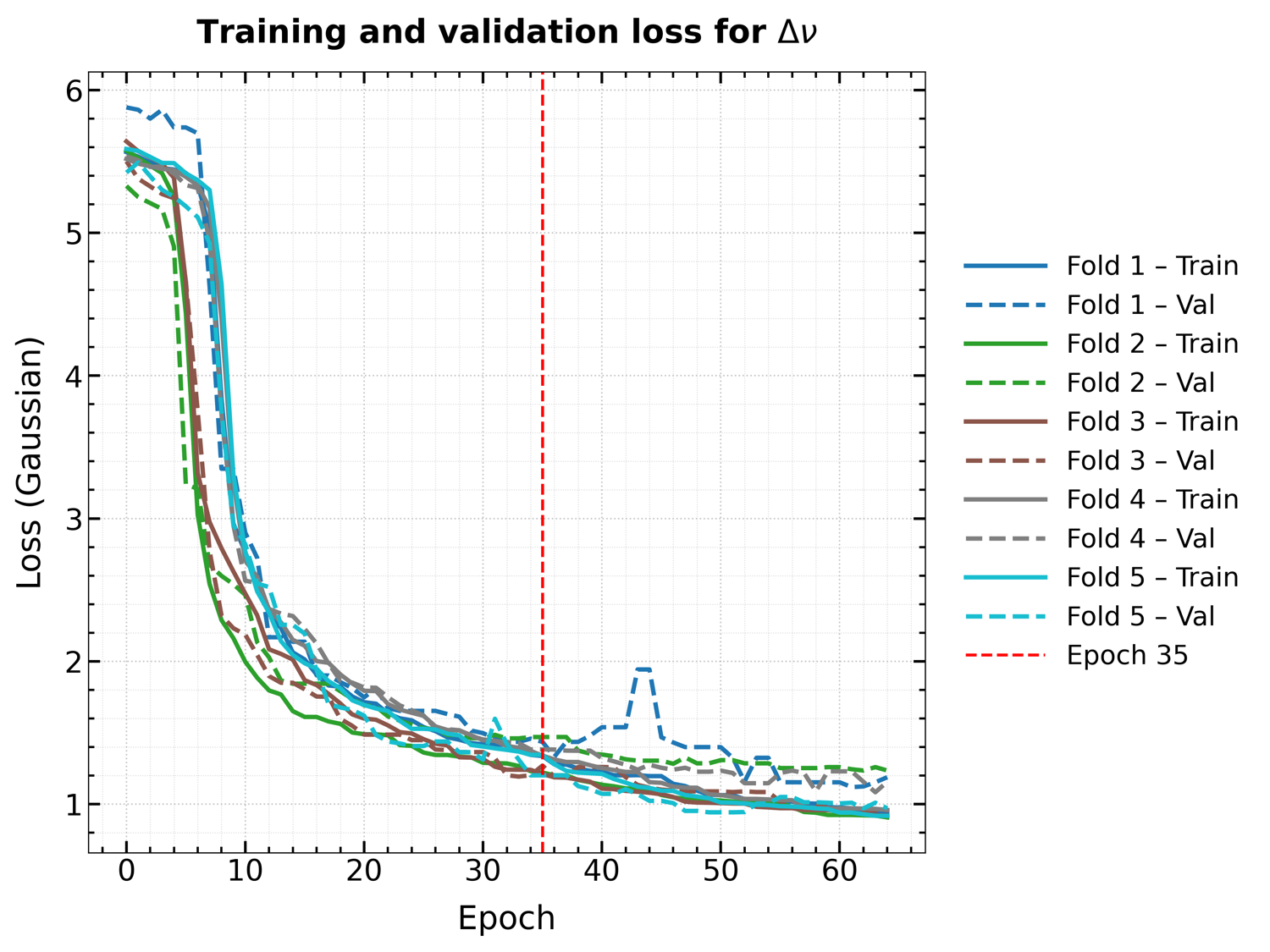}
        \caption{Loss evolution for $\Delta \nu$ using a Gaussian loss function (\ref{eq : gassian_loss}). The losses decrease sharply during the first few epochs, followed by a slower, consistent convergence trend. Occasional fluctuations in validation loss are observed at later epochs, reflecting fold-dependent variability, but overall stability is maintained.}
    \end{subfigure}
    \hfill
    \begin{subfigure}[t]{0.4\textwidth}
        \centering
        \includegraphics[width=\textwidth]{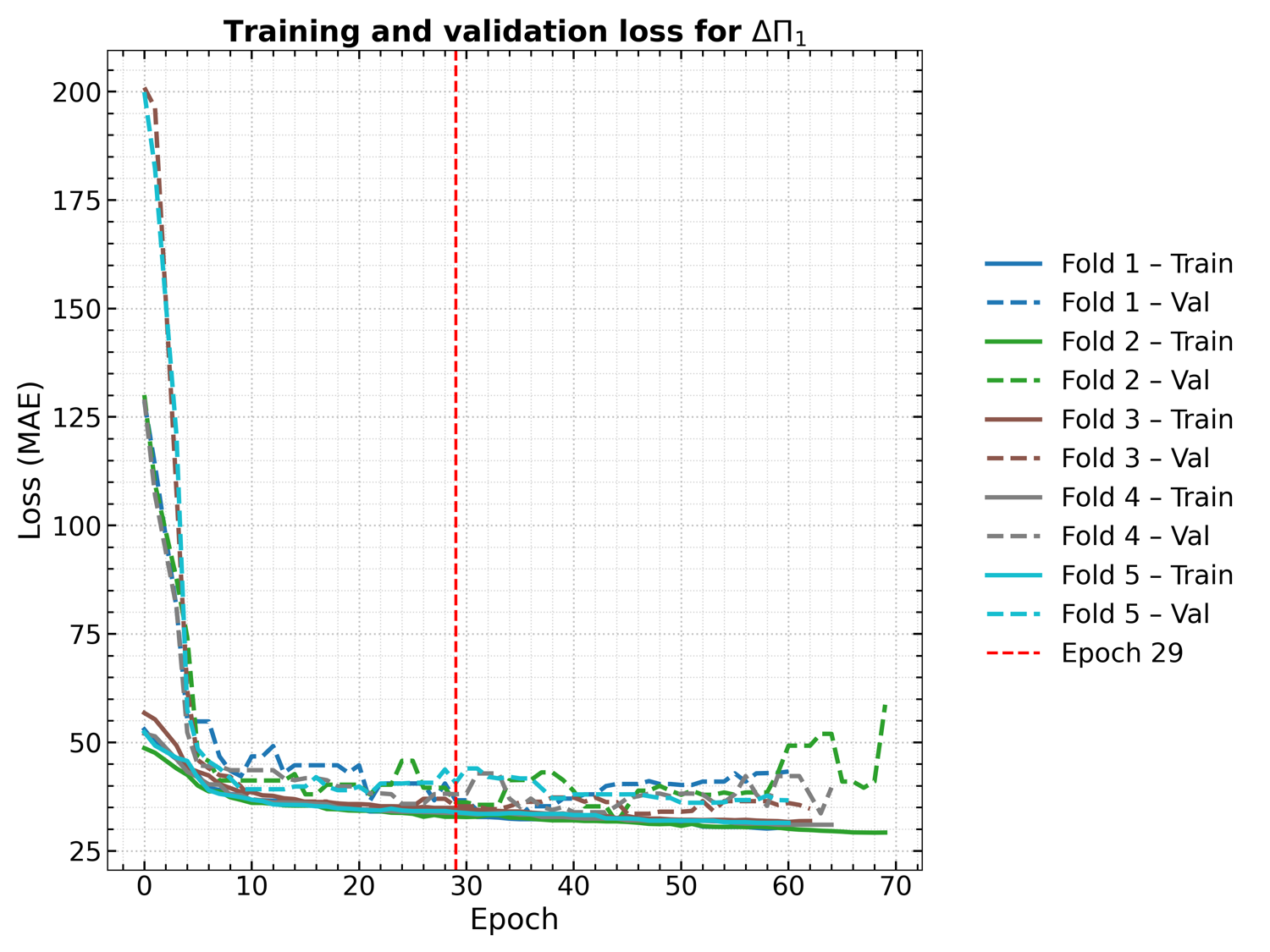}
        \caption{Loss evolution for $\Delta \Pi_{1}$ using mean absolute error (MAE). Similar to $\Delta \nu$, both training and validation losses exhibit a steep initial drop before stabilizing, with minor oscillations across folds toward the end of training.}
    \end{subfigure}

    \vskip 0.5cm

    \begin{subfigure}[t]{0.4\textwidth}
        \centering
        \includegraphics[width=\textwidth]{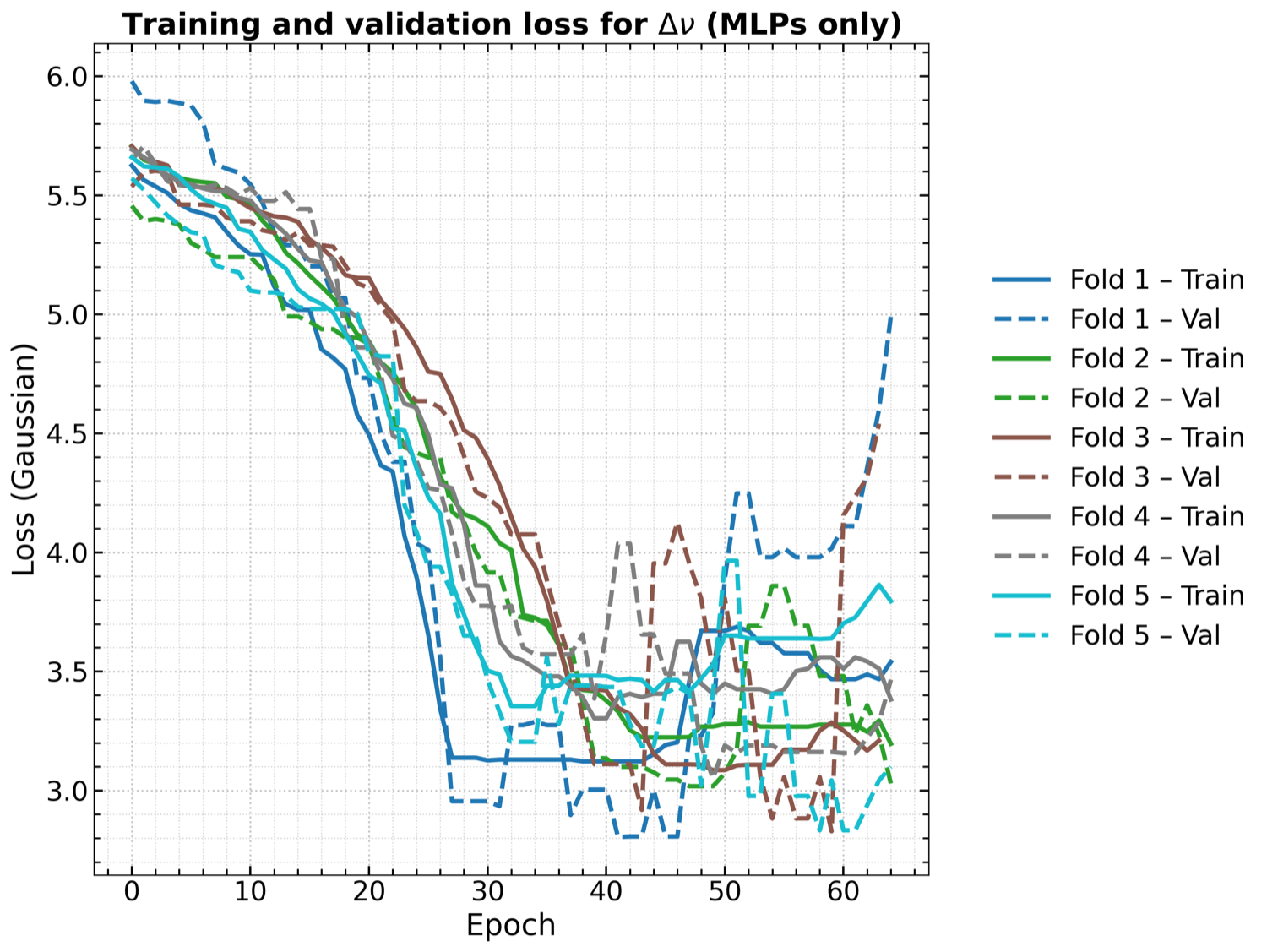}
        \caption{Training results using only the linear layers (loss function as Eq. (\ref{eq : gassian_loss})). The final, saturated loss is substantially higher than that obtained with the CNN--LSTM architecture, resulting in a 15\%--20\% overall decrease in \texttt{accuracy80} and \texttt{accuracy90} scores.}
    \end{subfigure}
    \hfill
    \begin{subfigure}[t]{0.4\textwidth}
        \centering
        \includegraphics[width=\textwidth]{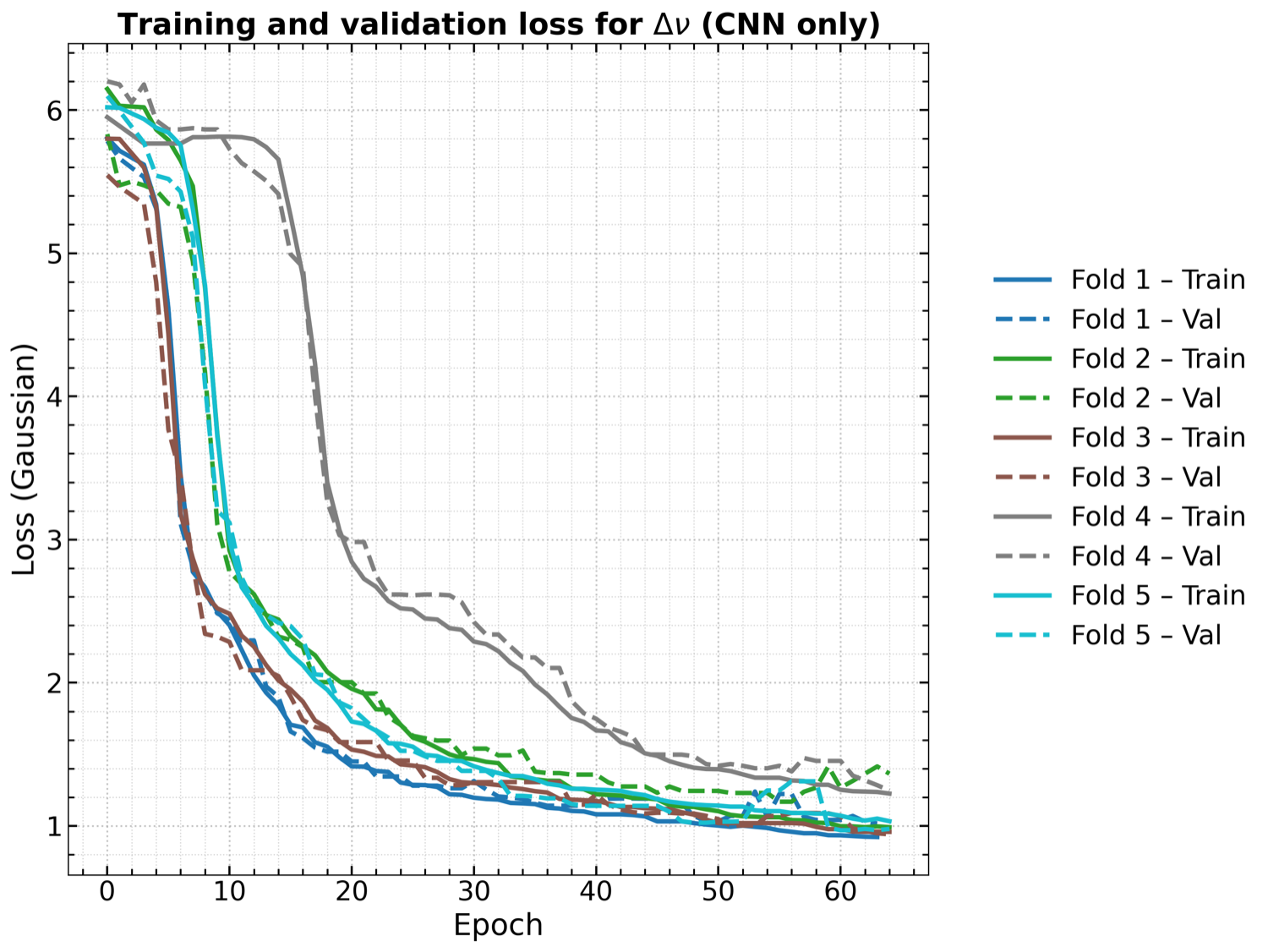}
        \caption{Training results using only the CNN layers. While the final, saturated loss is comparable to that obtained with the combined CNN--LSTM architecture, the generalization performance is significantly poorer, as evidenced by the large separation between the loss curves across different folds.}
    \end{subfigure}

    \caption{Cross-validation results comparing CNN--LSTM, linear-only, and CNN-only architectures. Training was deliberately extended over many epochs to ensure robust convergence. Final model selection was based on the epoch yielding the lowest \textit{average validation loss across all folds}, ensuring a fair estimate of generalization performance.}
    \label{fig:cross_val_res}
\end{figure*}

\section{Lightcurves of the Gaia stars in our 'horizontal' band region}

\begin{figure*}
    \centering
    \includegraphics[width=\textwidth]{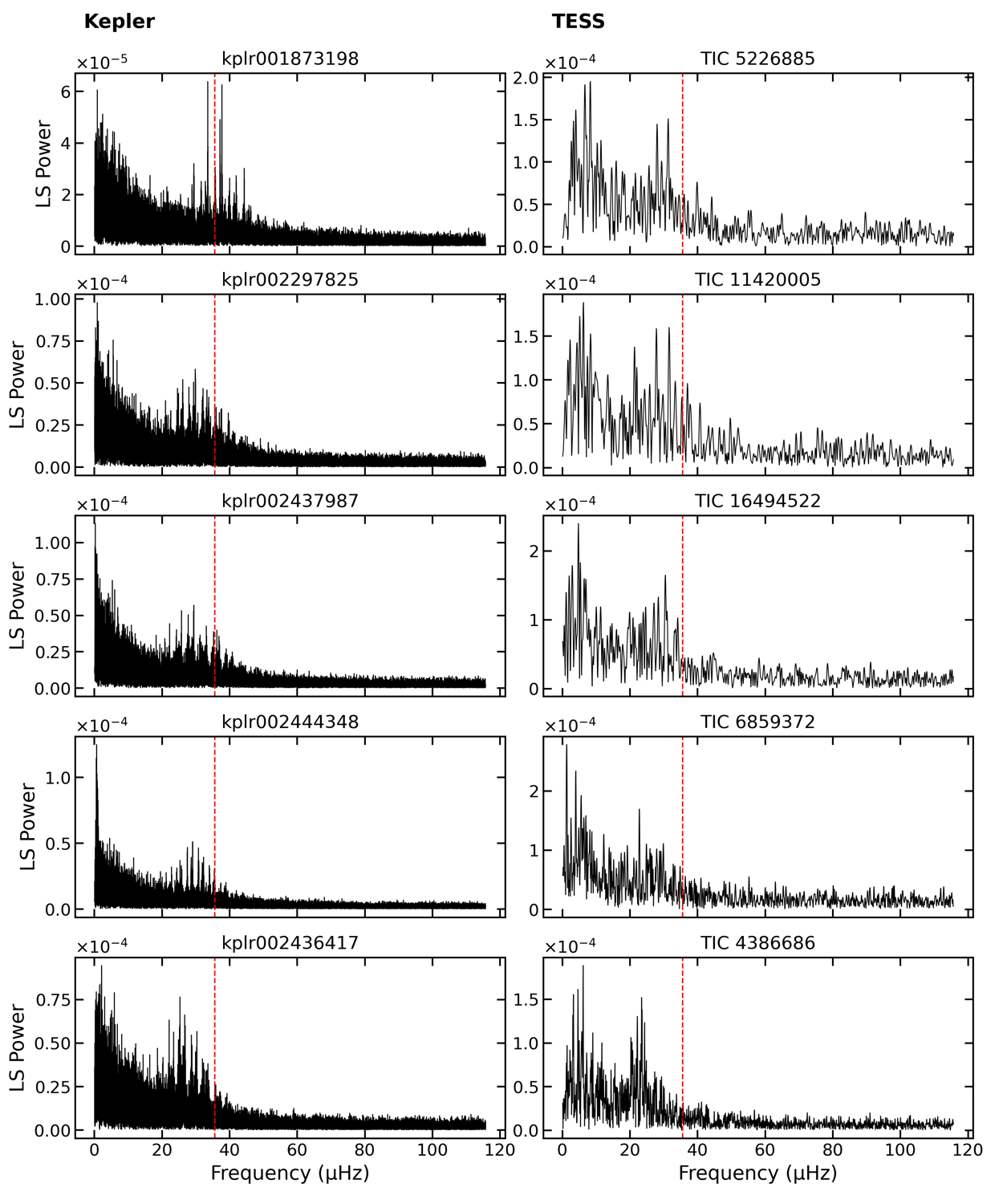}
    \caption{Power spectra of some red giants for which our Gaia XP inferences indicate $\Delta\nu \le 3.5~\mu\mathrm{Hz}$ and $\Delta\Pi_1 = 290 \pm 40~\mathrm{s}$. The dotted vertical red line indicates the upper limit of the permitted value of $\nu_{\mathrm{max}}$ corresponding to $\Delta\nu \le 3.5~\mu\mathrm{Hz}$.}
    \label{fig:power_spectra2}
\end{figure*}

\begin{figure*}
    \centering
    \includegraphics[width=\textwidth]{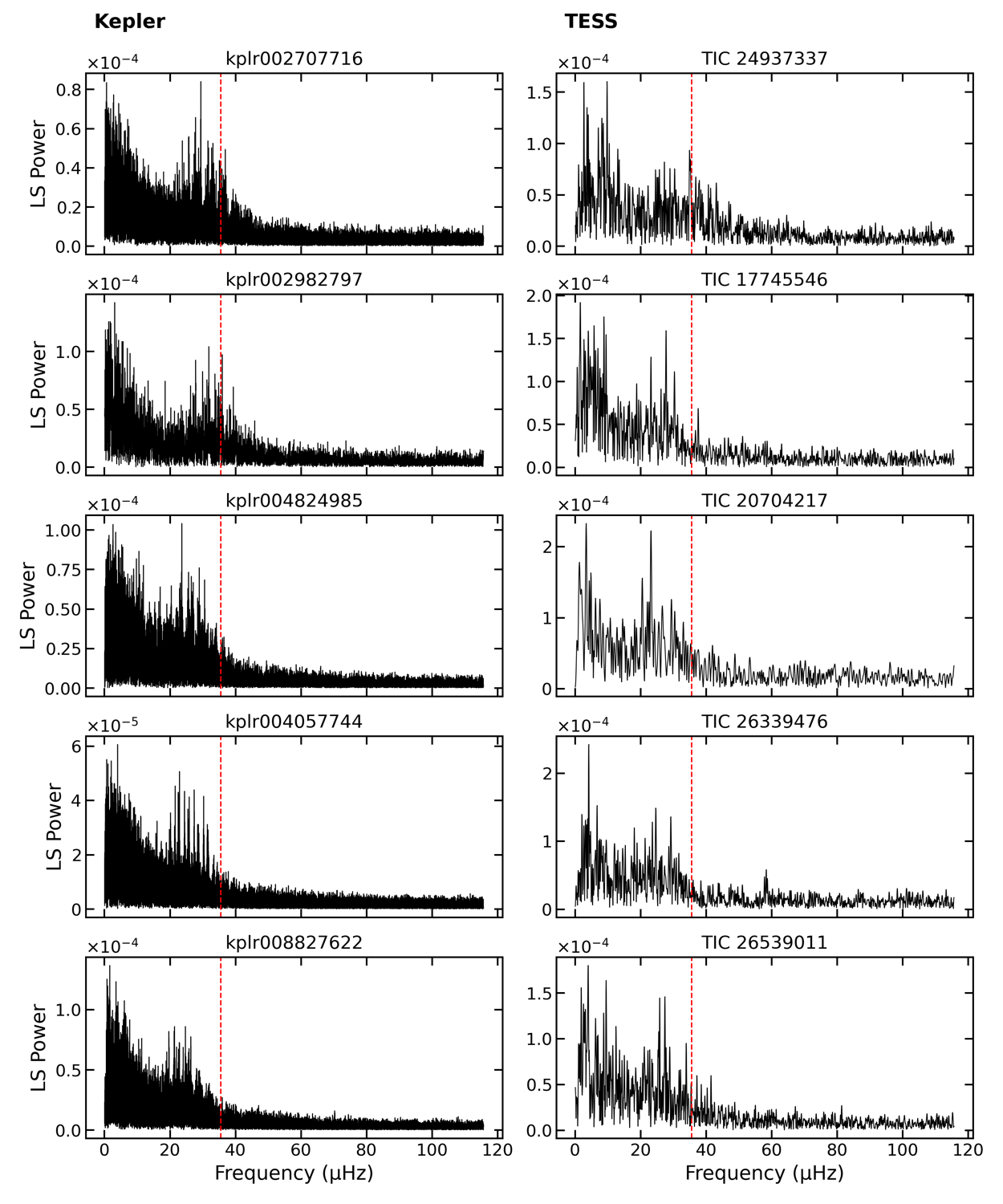}
    \caption{Power spectra of some red giants for which our Gaia XP inferences indicate $\Delta\nu \le 3.5~\mu\mathrm{Hz}$ and $\Delta\Pi_1 = 290 \pm 40~\mathrm{s}$. The dotted vertical red line indicates the upper limit of the permitted value of $\nu_{\mathrm{max}}$ corresponding to $\Delta\nu \le 3.5~\mu\mathrm{Hz}$.}
    \label{fig:power_spectra3}
\end{figure*}

\end{appendix}

%
%

\end{document}